\documentclass[%
  longbibliography,
  amsmath,amssymb,
  aps, reprint, secnumarabic, pre
]{revtex4-2}
\usepackage{tikz}
\usepackage{graphicx} 
\usepackage{subfigure}
\usepackage{multirow}
\usepackage{fancyhdr}
\usepackage{longtable}
\usepackage{parskip}
\usepackage[T1]{fontenc}
\usepackage{dcolumn} 
\usepackage{bm}       
\usepackage{amsfonts}  
\usepackage{amsmath}   
\usepackage{amssymb}   
\usepackage[]{hyperref}

\usepackage[utf8]{inputenc}
\usepackage[T1]{fontenc}
\usepackage{mathptmx}
\usepackage{etoolbox}
\usepackage{xcolor}

\usepackage{color}

\begin{document}
	
\title{Time-Delayed Dynamics in Regular Kuramoto Networks with Inertia: Multistability, Traveling Waves, Chimera States, and Transitions to Seizure-Like Activity}

\author{Esmaeil Mahdavi}
\affiliation{Department of Physics, Institute for Advanced Studies in Basic Sciences (IASBS), Zanjan 45137-66731, Iran}

\author{Philipp H\"ovel} 
\email{philipp.hoevel@uni-saarland.de}
\affiliation{Theoretical Physics and Center for Biophysics, Saarland University, Campus E2 6, Saarbrücken, 66123, Germany}

\author{Mina Zarei}
\affiliation{Department of Physics, Institute for Advanced Studies in Basic Sciences (IASBS), Zanjan 45137-66731, Iran}

\author{Farhad Shahbazi} 
\email{shahbazi@iut.ac.ir}
\affiliation{Department of Physics, Isfahan University of Technology, Isfahan 84156-83111, Iran}

\begin{abstract}  
	{This study examines the complex interplay between inertia and time delay in regular rotor networks within the framework of the second-order Kuramoto model. By combining analytical and numerical methods, we demonstrate that intrinsic time delays -- arising from finite information transmission speeds -- induce multistability among fully synchronized phase-locked states. Unlike systems without inertia, the presence of inertia destabilizes these phase-locked states, reduces their basin of attraction, and gives rise to nonlinear phase-locked dynamics over specific inertia ranges. In addition, we show that time delays promote the emergence of turbulent chimera states, while inertia enhances their spatial extent. Notably, the combined influence of inertia and time delay produces dynamic patterns reminiscent of partial epileptic seizures. These findings provide new insights into synchronization phenomena by revealing how inertia and time delay fundamentally reshape the stability and dynamics of regular rotor networks, with broader implications for neuronal modeling and other complex systems.}
\end{abstract}

\maketitle

\section{Introduction}

Synchronization is a fundamental phenomenon in both natural and engineered systems, where individual components collectively exhibit coherent dynamics. Since its introduction in 1975, the Kuramoto model has served as a cornerstone for studying synchronization, modeling interactions through first-order nonlinear coupled oscillators~\cite{kuramoto1975self}. This framework has provided key insights into synchronization in chemical, biological, and mechanical systems~\cite{tyson1973some,mirollo1990synchronization,nijmeijer2003synchronization}. However, the first-order Kuramoto model has notable limitations: it often converges rapidly to partially synchronized states and fails to capture frequency adaptation observed in certain biological systems, such as firefly populations~\cite{ermentrout1991adaptive}. To address these shortcomings, Ermentrout introduced an extension incorporating inertia, giving rise to the second-order Kuramoto model~\cite{ermentrout1991adaptive,tanaka1997self}. This modification fundamentally changes the character of phase transitions~\cite{tanaka1997first} and provides a more realistic framework for analyzing complex systems, including Josephson junction arrays, power grids, and neuronal networks~\cite{trees2005synchronization,filatrella2008analysis,rohden2012self,rohden2014impact,grzybowski2016synchronization}. The second-order Kuramoto model is especially relevant for neuroscience, where simplified neuronal prototypes can be expressed as second-order nonlinear differential equations~\cite{dolan2005phase,majtanik2006desynchronization,sakyte2011self}. Furthermore, Josephson junction dynamics have been shown to closely mimic neuronal behavior~\cite{mishra2021neuron}.
Another critical factor influencing synchronization is time delay, which arises naturally from finite signal transmission speeds in systems such as neural networks, electronic circuits, and laser arrays~\cite{kozyreff2000global,reddy2000experimental,yeung1999time,kerszberg1990synchronization,waibel2013phoneme}. Time delays can profoundly affect system dynamics by inducing multistability~\cite{schuster1989mutual,mahdavi2025synchronization,yeung1999time,choi2000synchronization,madadi2018delay}, chimera states, and first-order phase transitions~\cite{ameli2021time}. In neuronal networks, time delays have been shown to foster synchronization between distinct populations~\cite{vicente2008dynamical} or even trigger transitions from synchronized to desynchronized states~\cite{lehnert2011loss}.

Incorporating time delay into the Kuramoto model with inertia substantially increases the complexity of network dynamics. In binary rotor systems, time delay induces multistability, whereas greater inertia reduces the stability of phase-locked states. Their combined influence can generate periodic or chaotic non-phase-locked solutions, with time delay itself acting as a control parameter for chaos~\cite{mahdavi2025synchronization}. In globally coupled rotor networks, the interplay between inertia and time delay gives rise to both subcritical (first-order) and supercritical (second-order) bifurcations between synchronized and incoherent states~\cite{metivier2020onset}. Analytical investigations of the second-order Kuramoto model with delay have further explored these phenomena~\cite{mahdavi2025synchronization,metivier2020onset,hong2002spontaneous,prousalis2022synchronization,dai2018interplay}.

The interaction between inertia and time delay adds significant complexity to network dynamics. 
Previous studies have shown that time delay can induce multistability in binary rotor systems and regular networks, as well as alter the stability of phase-locked states~\cite{schuster1989mutual,ameli2025synchronization}.
Together, inertia and delay can generate periodic or chaotic behavior~\cite{mahdavi2025synchronization} and modulate phase transitions between synchronized and incoherent states. Although the influence of time delay has been widely investigated in globally coupled networks and first-order models, its role in regular networks with local interactions, and within the second-order Kuramoto framework remains insufficiently understood.

This work addresses this gap by examining the joint effects of inertia and time delay on synchronization dynamics in regular rotor networks governed by the second-order Kuramoto model. Using analytical derivations and numerical simulations, we investigate their impact on multistability, phase-locked states, and chimera patterns. Our results provide novel insights into the stability boundaries of synchronized states and reveal dynamic regimes reminiscent of partial epileptic seizures. These findings advance the understanding of synchronization in complex systems and offer implications for neuronal modeling and related applications.
 
The paper is structured as follows: Section~\ref{model} introduces the mathematical model and methods used to analyze fully synchronized phase-locked solutions and their stability. Section~\ref{results} presents numerical results that illustrate the system's dynamics, and Section~\ref{conclusion} concludes with a summary of the main findings and their broader implications.

\section{Model an Method\label{model}}
We use the second-order Kuramoto model to study the effects of time-delayed interactions in a network of coupled oscillators. For a system of $N$ identical rotors, the model is given by~\cite{tanaka1997self}:
\begin{equation}
m\ddot{\theta}_{i}(t)+\dot{\theta}_{i}(t)=\omega+\frac{\alpha}{k_i} \sum_{j=1}^{N}a_{ij}\sin(\theta_j(t-\tau)-\theta_i(t)),
\label{Eq:maineq}
\end{equation}
where $\theta_{i}$ denotes the phase of the $i$th rotor, $m$ the inertia, $\omega$ the intrinsic frequency, $\alpha$ the coupling strength, $\tau$ the time delay, and $k_i$ the degree of node $i$. 

The adjacency matrix $a_{ij}$ specifies the network connectivity, with $a_{ij}=1$ indicating a link between rotors $i$ and $j$, and $a_{ij}=0$ otherwise. Throughout this study, we consider a regular ring topology with $P$ neighbors to either side of a node and use the following parameters: $N=1000$, $P=5$, \textit{i.e.}, $k_i=\langle k \rangle=2P=10$ for $i=1,\dots,N$, $\omega=1$, and  $\alpha=1$.

To analyze the stability of the fully synchronized state and determine the corresponding phase-locked frequency $\Omega_{f}$, we derive the master stability function. In this state, all rotors share a common temporal phase evolution~\cite{mahdavi2025synchronization}:
\begin{equation}
{\theta}_{i}(t)=\Omega_{f} t+\beta, 
\label{Eq:phaselocked1}
\end{equation}
where $\beta$ is a constant phase offset. To assess the stability of this solution, we introduce a small perturbation $\xi_i$ to the rotor phases and linearize the system, obtaining:
\begin{equation}
m\ddot{\boldsymbol{\xi}}+\dot{\boldsymbol{\xi}}={\alpha}\cos({\Omega_f}\tau) (\boldsymbol{\xi}^{\tau}-\boldsymbol{\xi})-\frac{{\alpha}}{\langle k \rangle}\cos({\Omega_f}\tau) \boldsymbol{L}\boldsymbol{\xi}^{\tau}.
\label{Eq:xi}
\end{equation}
where $\boldsymbol{\xi}(t) = [\xi_1(t), \dots, \xi_N(t)]^T$ and $\boldsymbol{\xi}^\tau(t) = [\xi_1(t-\tau), \dots, \xi_N(t-\tau)]^T$ denote the perturbation vectors at times $t$ and $t-\tau$, respectively.

Here, $\boldsymbol{L}$ denotes the Laplacian matrix with elements $l_{ij}=\delta_{ij}k_{i}-a_{ij}$. Diagonalizing $\boldsymbol{L}$ yields
\begin{equation}
m\ddot{\boldsymbol{\eta}}+\dot{\boldsymbol{\eta}}={\alpha}\cos({\Omega_f}\tau) (\boldsymbol{\eta}^{\tau}-\boldsymbol{\eta})-\frac{{\alpha}}{\langle k \rangle}\cos({\Omega_f}\tau) \boldsymbol{\Lambda}\boldsymbol{\eta}^{\tau},
\label{Eq:xi2}
\end{equation}
where $\boldsymbol{\Lambda}$ is a diagonal matrix containing the Laplacian eigenvalues $\lambda_i$. Assuming solutions of the form $\eta_i(t) \propto \exp(z_i t)$ leads to the characteristic equation governing the stability of each node:
\begin{equation}
\begin{aligned}
m z^2_i+z_i &={\alpha}\cos({\Omega_f}\tau) e^{-z_i\tau} \left(1- e^{z_i\tau} -\frac{\lambda_i}{\langle k \rangle}\right).
\end{aligned}
\label{Eq:xi3}
\end{equation}
Equation~\eqref{Eq:xi3} constitutes the master stability function of the system. The phase-locked state is stable if all $z_i$ have negative real parts; otherwise, it is unstable.

The phase-locked frequency $\Omega_f$ in the fully synchronized state is obtained from the roots of the transcendental equation
\begin{equation}
\Omega_f-\omega+{\alpha}\sin(\Omega_f\tau)=0.
\label{Eq:root}
\end{equation}

The degree of synchrony in the network is characterized by the Kuramoto order parameter $r(t)$, defined as
\begin{eqnarray}
	r (t)& = & \frac{1}{N} \left|\sum_{j=1}^N e^{i\theta_{j}(t)}\right|,
	\label{Eq:r}
\end{eqnarray} 
where $0 \leq r(t) \leq 1$, with $r(t)=1$ denoting perfect synchrony and $r(t)=0$ indicating complete incoherence.

In addition, the local order parameter $R_l(t)$ quantifies the synchrony between rotor $l$ and its $2M$ nearest neighbors:
\begin{equation}
R_{l}(t)=\frac{1}{2M+1}\left|\sum_{j=l-M}^{l+M}e^{i\theta_j(t)}\right|,
\label{Eq:rlocal}
\end{equation}
with $0 \leq R_l(t) \leq 1$. Here, $R_l(t)=1$ indicates perfect local synchrony, while $R_i(t)=0$ denotes complete incoherence between rotor $i$ and its neighborhood. In the following, we use $M=25$, that is, we calculate the local order via an average over $2M=50$ node.

To characterize the spatial organization of phase relationships within the network, we use the cosine similarity matrix $S$ given by
\begin{eqnarray}
S_{ij} & = &\lim_{	\Delta t \to \infty} \frac{1}{\Delta t} \int_{t_{s}}^{t_{s}+\Delta t} \cos(\theta_{i}(t)-\theta_{j}(t)) \;\mathrm{d}t,
\label{Eq:dij}
\end{eqnarray}
where $-1 \leq S_{ij} \leq 1$, and $t_s$ denotes the transient time. A value of $S_{ij}=1$ indicates that rotors $i$ and $j$ are perfectly in phase, whereas $S_{ij}=-1$ corresponds to a perfectly anti-phase relationship. For the temporal similarity and periodicity, we also calculate the average autocorrelation function $C$ defined as 
\begin{equation}
C(\tau') = \left\langle\left| \frac{1}{N}\sum_{j=1}^N e^{i\theta_{j}(t)} e^{-i\theta_{j}(t-\tau')}\right|\right\rangle.
\end{equation}

Furthermore, we classify the observed chimera states following the methodology of Kemeth et al.~\cite{kemeth2016classification}, which combines a spatial coherence function with a measure of temporal correlation. The spatial coherence function is obtained by first evaluating the local curvature of the spatial phase profile using a discrete Laplacian operator. Specifically, the discrete Laplacian $\mathbf{D}$ is rescaled and applied to the spatial phase data $f$ at time $t$:
\begin{equation}
\hat{\mathbf{D}}f=\Delta x^2 \mathbf{D}f=f(x+\Delta x,t)-2f(x,t)+f(x-\Delta x,t),
\label{Eq:df}
\end{equation}
where the phases are represented as complex numbers on the unit circle. The maximum value of $|\hat{\mathbf{D}}f|$ is $D_m = 4$, and the spatial coherence function $g_0(t)$ is subsequently computed as:
\begin{equation}
	g_0(t) = \int_{0}^{0.01 D_m} g(t,|\hat{\mathbf{D}}f|)  d|\hat{\mathbf{D}}f|,
	\label{Eq:gt}
\end{equation}
denotes the normalized probability density function of $|\hat{\mathbf{D}}f|$ at time $t$. A value of $g_0(t)=1$ corresponds to a fully synchronized state, $g_0(t)=0$ indicates complete incoherence, and intermediate values $0 < g_0(t) < 1$ signify the coexistence of coherent and incoherent regions.

To quantify temporal correlations in the network, we calculate the pairwise correlation coefficients $\rho_{ij}$ between the phase time series of different rotors. For the phase time series $\theta_i(t)$ of the $i^{\text{th}}$ rotor, with mean $\mu_i$ and standard deviation $\sigma_i$, the pairwise correlation coefficient is defined as
\begin{equation}
{\rho}_{ij}=\frac{\langle (\theta_i-\mu_i)^* (\theta_j-\mu_j) \rangle}{\sigma_i \sigma_j},
\label{Eq:rho}
\end{equation}
where $\langle \cdot \rangle$ and $^*$ denote the temporal mean and complex conjugate, respectively. The normalized distribution function $h$ of the elements of the correlation matrix $\hat{\mathbf{R}}=\{|{\rho}_{ij}|\}, i\neq j,$ is then used to compute the fraction $h_0$ of pairs with strong temporal correlation
\begin{equation}
h_0=\sqrt{\int_{0.99}^{1} h(|\rho|) d|\rho|}.
\label{h0}
\end{equation} 

Table \ref{tab:cond-chimera} summarizes the classification criteria for different types of chimera states based on the values of the spatial coherence function, $g_0(t)$, and the temporal correlation measure, $h_0$.  

\begin{table}[t!]
\centering
\caption{Characterization of chimera states based on $g_0(t)$ and $h_0$.}
  \begin{tabular}{ |l|l|}
    \hline
    Classification & Conditions of functions $g_0(t)$ and $h_0$ \\ \hline\hline
    	static stationary& $g_0(t)=$ constant,  $h_0>0$, $\{\forall t: 0<g_0(t)<1\}$ \\ \hline
	moving stationary& $g_0(t)=$ constant,  $h_0\approx0$, $\{\forall t: 0<g_0(t)<1\}$ \\ \hline
	static turbulent& $g_0(t)=$ irregular,  $h_0>0$, $\{\forall t: 0<g_0(t)<1\}$ \\ \hline
	moving turbulent& $g_0(t)=$ irregular,  $h_0\approx0$, $\{\forall t: 0<g_0(t)<1\}$  \\ \hline
	static breathing& $g_0(t)=$ oscillatory,\\
	&$h_0>0$, $\{\forall t: 0<g_0(t)<1\}$ \\ \hline
	moving breathing& $g_0(t)=$ oscillatory, \\
	&$h_0\approx0$, $\{\forall t: 0<g_0(t)<1\}$  \\ \hline
	transient& $\{\exists t_0: g_0(t)=0 \lor g_0(t)=1\}$\\
	&$ \land \{\forall t<t_0: 0<g_0(t)<1\}$ \\ \hline
	no chimera & $\exists t: g_0(t)=0 \lor g_0(t)=1$ \\
    \hline
  \end{tabular}
\label{tab:cond-chimera}
\end{table}

\section{Results and Discussion\label{results}}

\begin{figure*}[th!]
	\centering
	\renewcommand\thesubfigure{\fontsize{10}{10}\selectfont (\alph{subfigure})}
	\fbox{\subfigure[\label{fig1:a}]{\includegraphics[width=2.0\columnwidth]{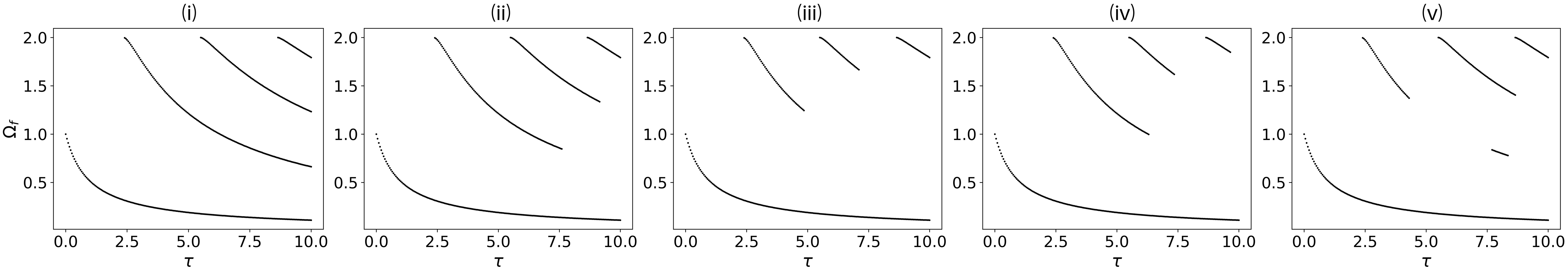}}}
	\fbox{\subfigure[\label{fig1:b}]{\includegraphics[width=2.0\columnwidth]{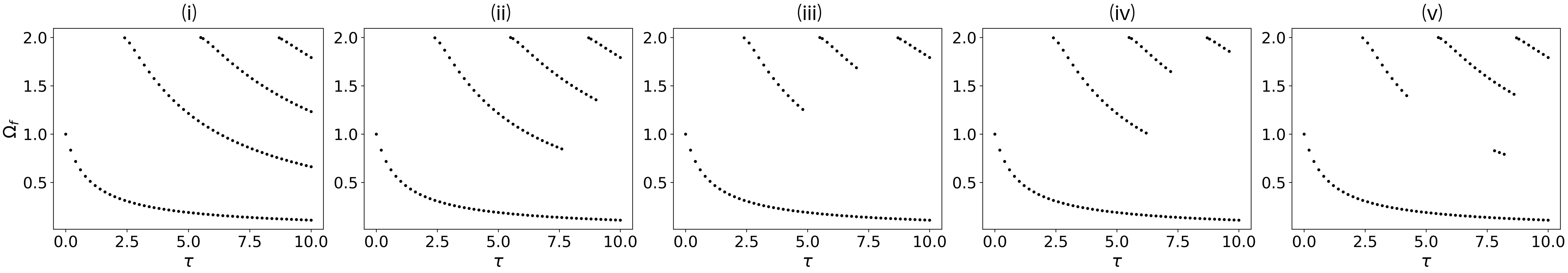}}}	
	\caption{(a) Analytically stable, and (b) numerically linear phase-locked frequencies $\Omega_f$ corresponding to complete synchrony versus time delay. The numerical results are presented in the forward direction, using parameter continuation initialized from the previous state. Different panels correspond to inertia values: (i) $m=0$, (ii) $m=0.6$, (iii) $m=1$, (iv) $m=5$, and (v) $m=10$. Other parameters: $N=1000$, $P=5$, \textit{i.e.}, $k_i=\langle k \rangle=2P=10$ for $i=1,\dots,N$, $\omega=1$, and  $\alpha=1$.}
	\label{fig1:w-tau}
\end{figure*}
\begin{figure*}[th!]
	\centering
	\renewcommand\thesubfigure{\fontsize{10}{10}\selectfont (\alph{subfigure})}
	\fbox{\subfigure[\label{fig2:a}]{\includegraphics[width=2.0\columnwidth]{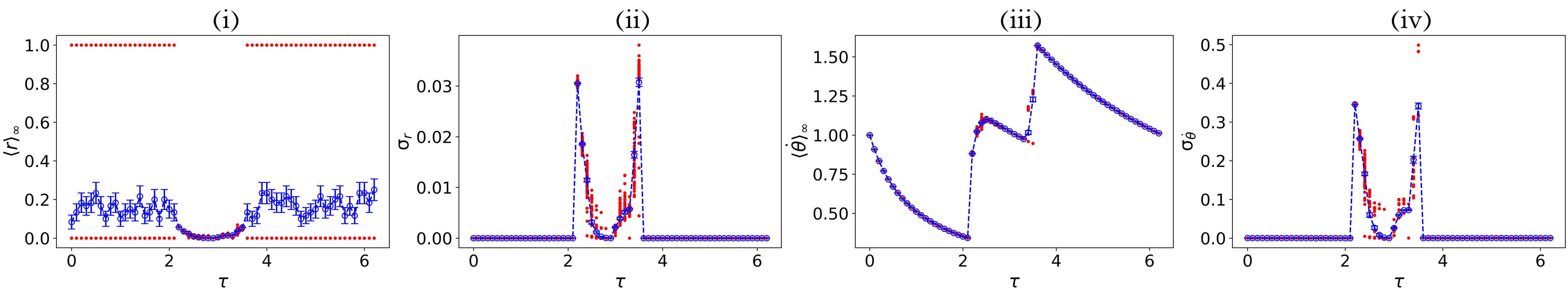}}}
	\fbox{\subfigure[\label{fig2:b}]{\includegraphics[width=2.0\columnwidth]{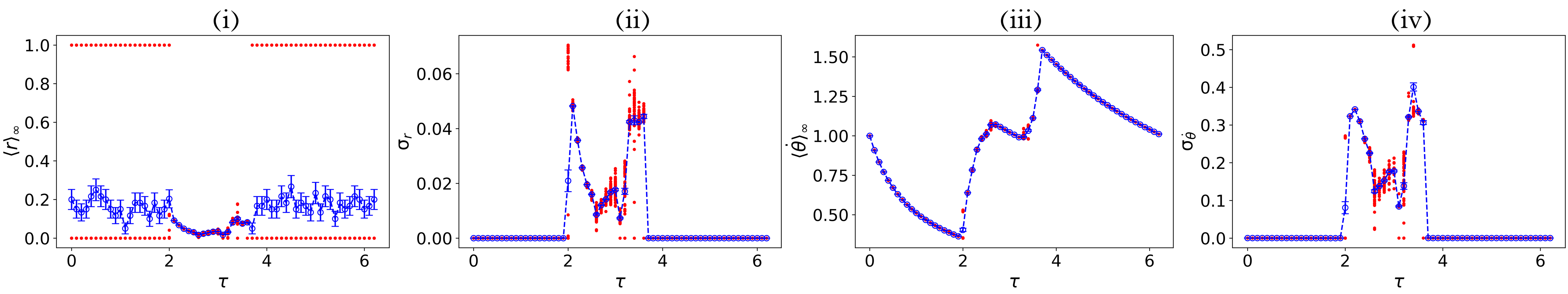}}}
	\fbox{\subfigure[\label{fig2:c}]{\includegraphics[width=2.0\columnwidth]{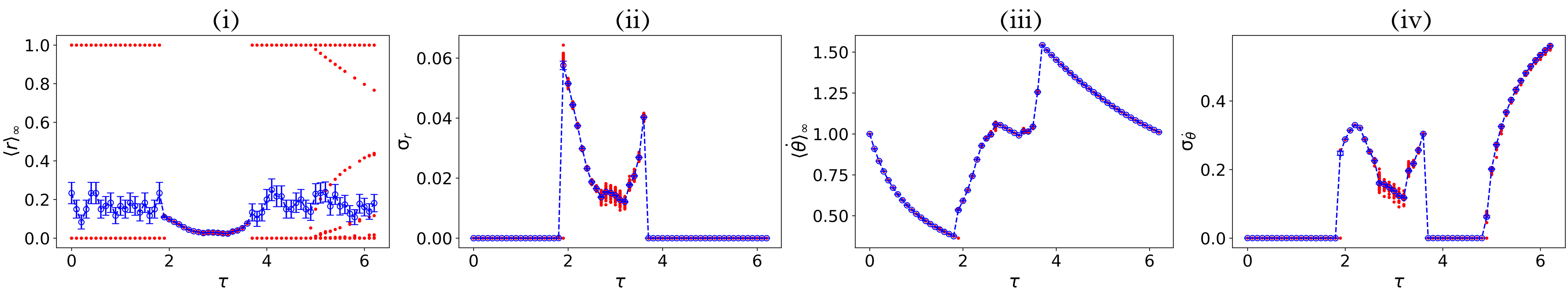}}}
	\fbox{\subfigure[\label{fig2:d}]{\includegraphics[width=2.0\columnwidth]{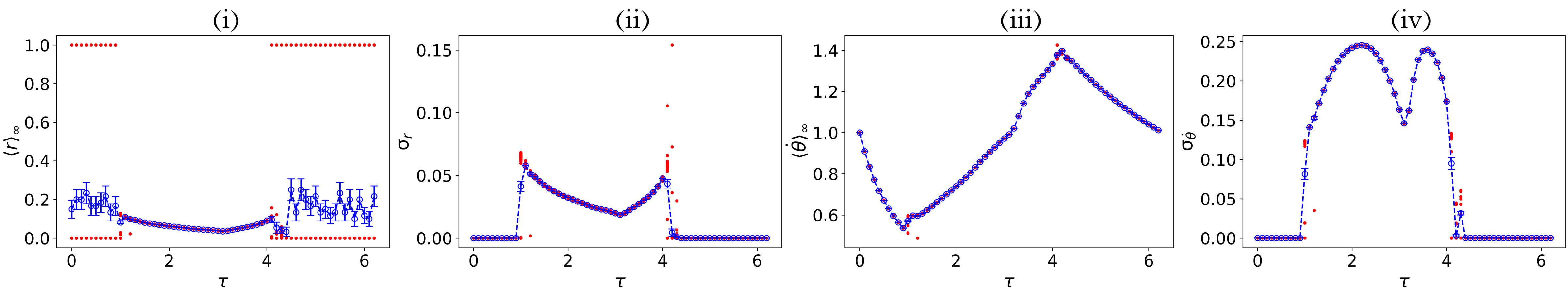}}}	
	\caption{(i) Time-averaged order parameter in the steady state, (ii) standard deviation of order parameter fluctuations, (iii) mean rotor angular velocity, and (iv) standard deviation of the angular velocity as functions of time delay. Each row corresponds to a different inertia value: (a) $m = 0$, (b) $m = 0.6$, (c) $m = 1$, and (d) $m = 5$. Red dots represent scatter plots of results from 60 distinct initial conditions, while blue dots indicate ensemble averages across these initial conditions, accompanied by standard error bars. Other parameters as in Fig.~\ref{fig1:w-tau}.}
	\label{fig2:r-tau0}
\end{figure*}
\begin{figure*}[th!]
	\centering
	\subfigure[\label{fig3:a}]{\includegraphics[width=0.51\columnwidth]{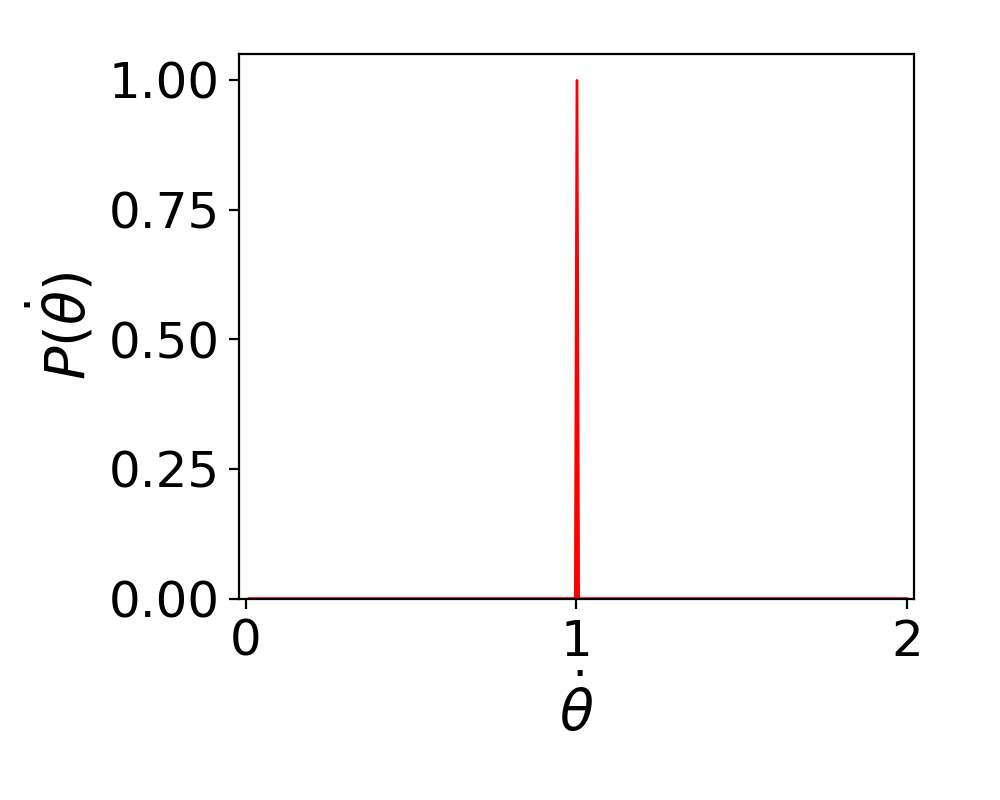}}
	\subfigure[\label{fig3:b}]{\includegraphics[width=0.51\columnwidth]{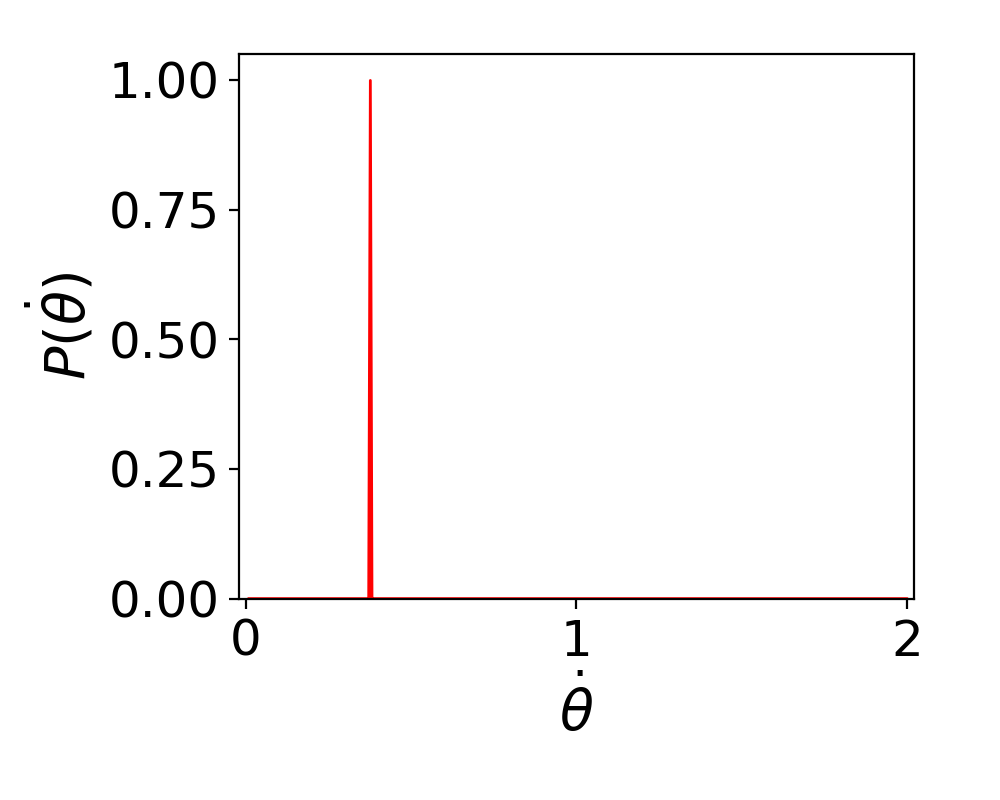}}
	\subfigure[\label{fig3:c}]{\includegraphics[width=0.51\columnwidth]{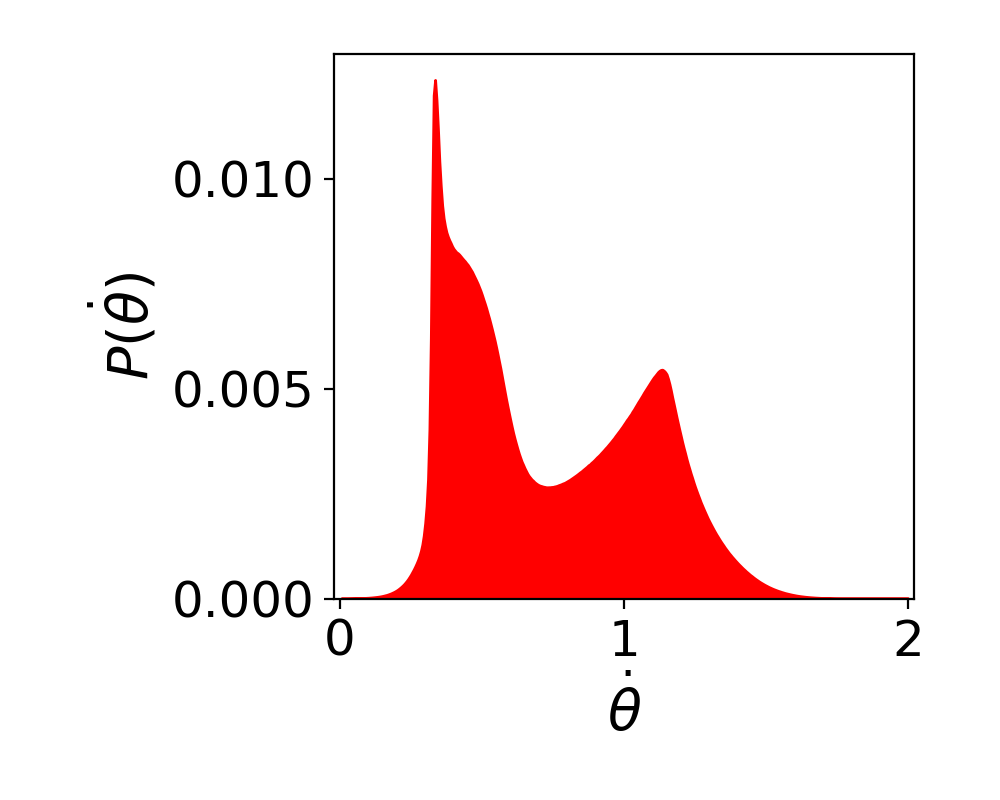}}
	\subfigure[\label{fig3:d}]{\includegraphics[width=0.51\columnwidth]{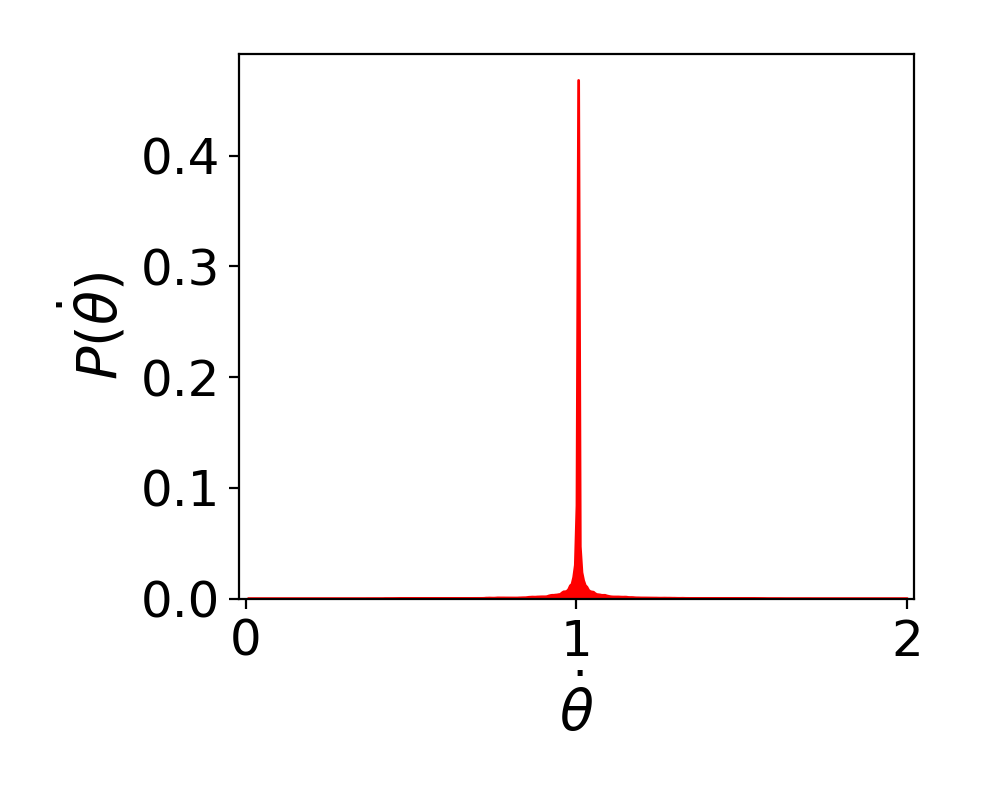}}
	\subfigure[\label{fig3:e}]{\includegraphics[width=0.51\columnwidth]{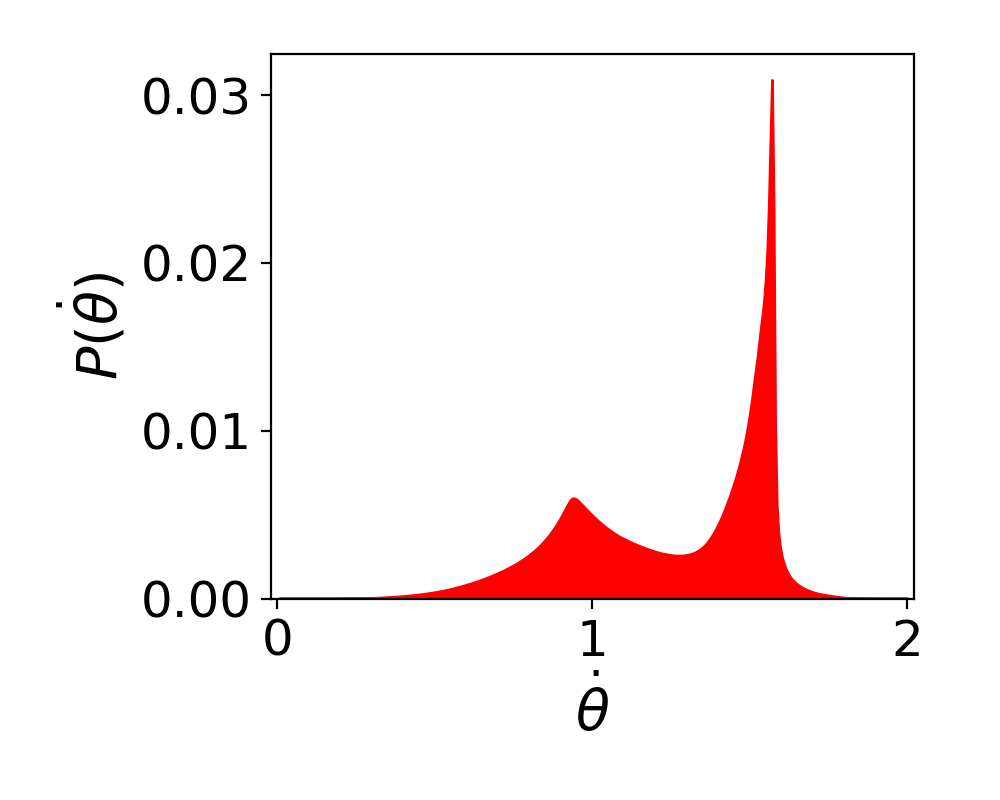}}
	\subfigure[\label{fig3:f}]{\includegraphics[width=0.51\columnwidth]{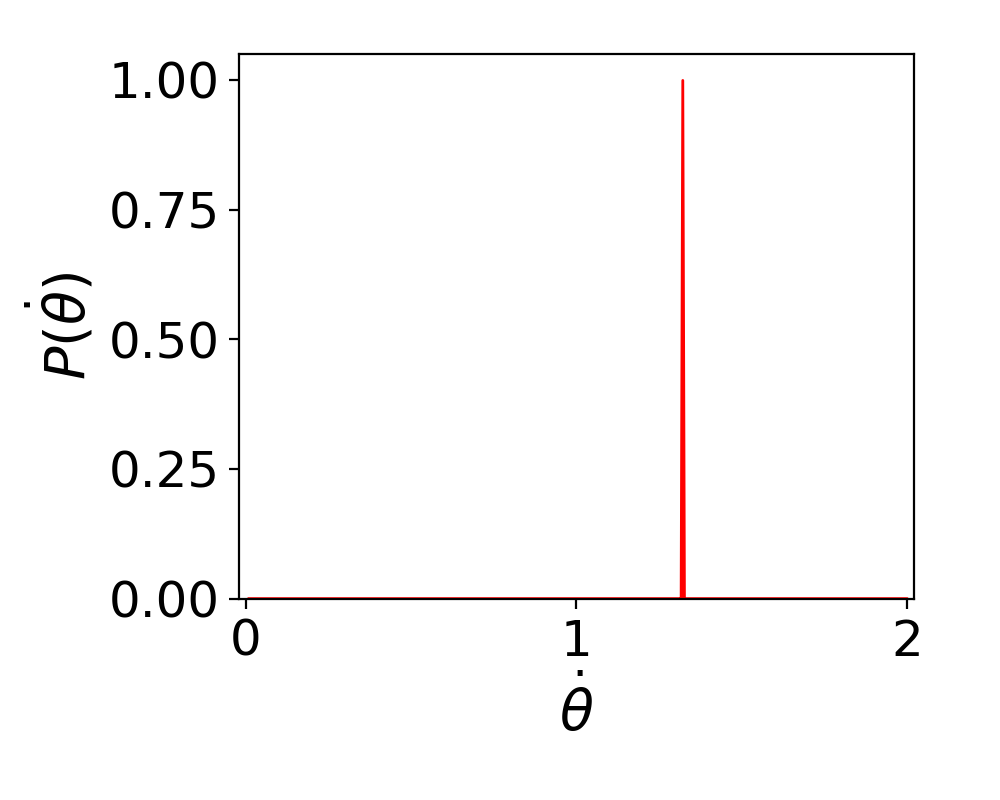}}
	\subfigure[\label{fig3:g}]{\includegraphics[width=0.51\columnwidth]{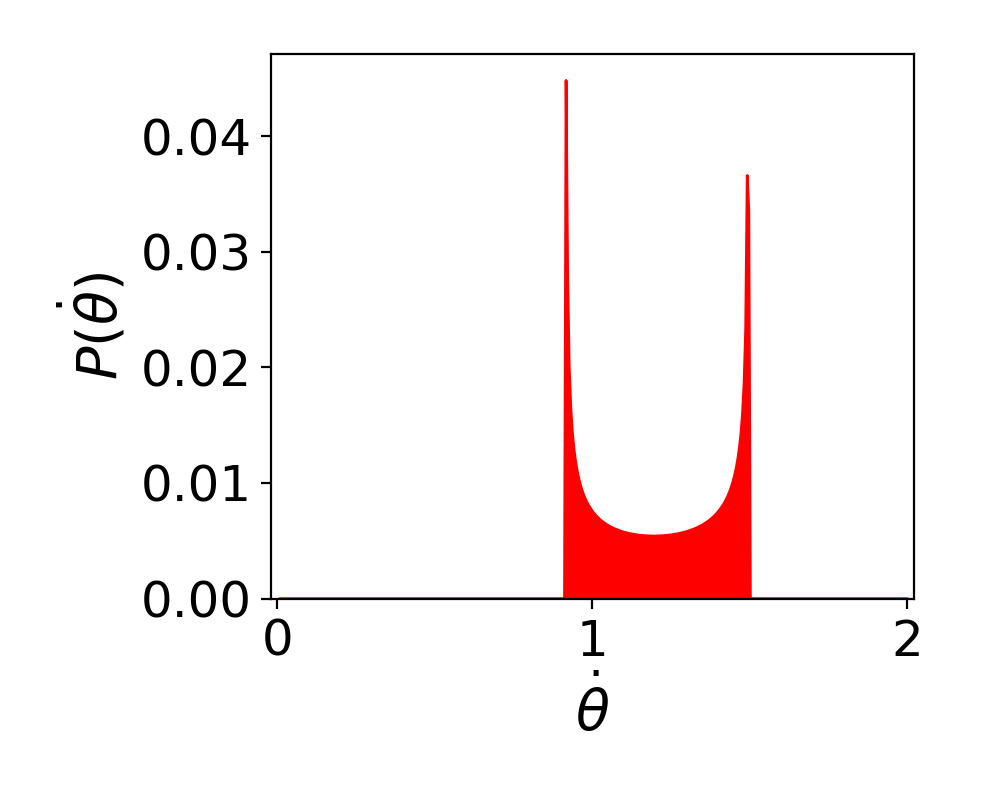}}
	\subfigure[\label{fig3:h}]{\includegraphics[width=0.51\columnwidth]{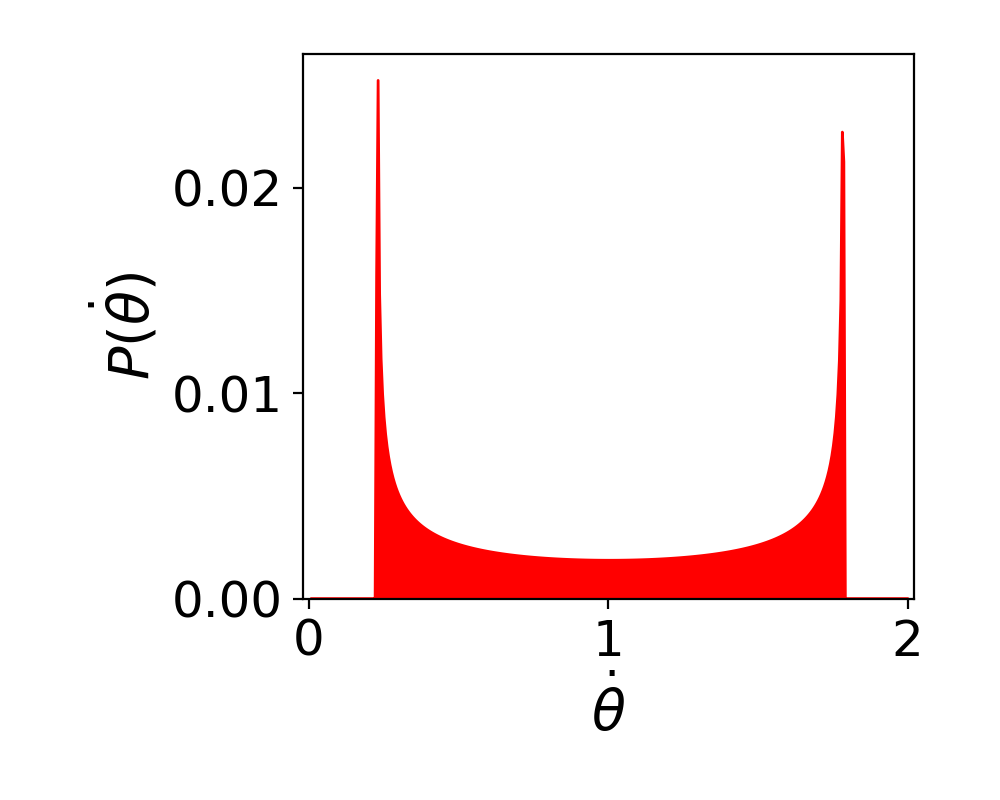}}
	\caption{The probability density function of angular velocities for an inertia value $m=1$. Each panel corresponds to different time delays: (a) $\tau=0$, (b) $\tau=1.8$, (c) $\tau=2.2$, (d) $\tau=3.1$, (e) $\tau=3.6$, (f) $\tau=4.5$, (g) $\tau=5$, and (h) $\tau=6.2$. Other parameters as in Fig.~\ref{fig1:w-tau}.}
	\label{fig3:pwm1}
\end{figure*}
\begin{figure*}[th!]
	\centering
	\subfigure[\label{fig4:a}]{\includegraphics[width=0.51\columnwidth]{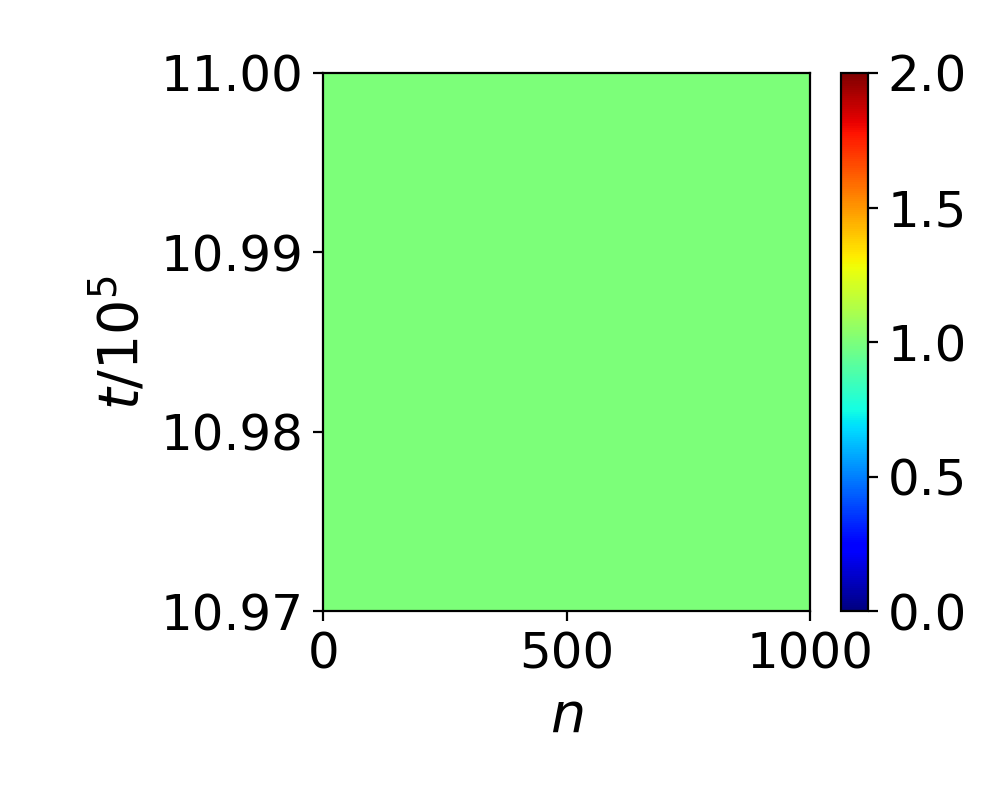}}
	\subfigure[\label{fig4:b}]{\includegraphics[width=0.51\columnwidth]{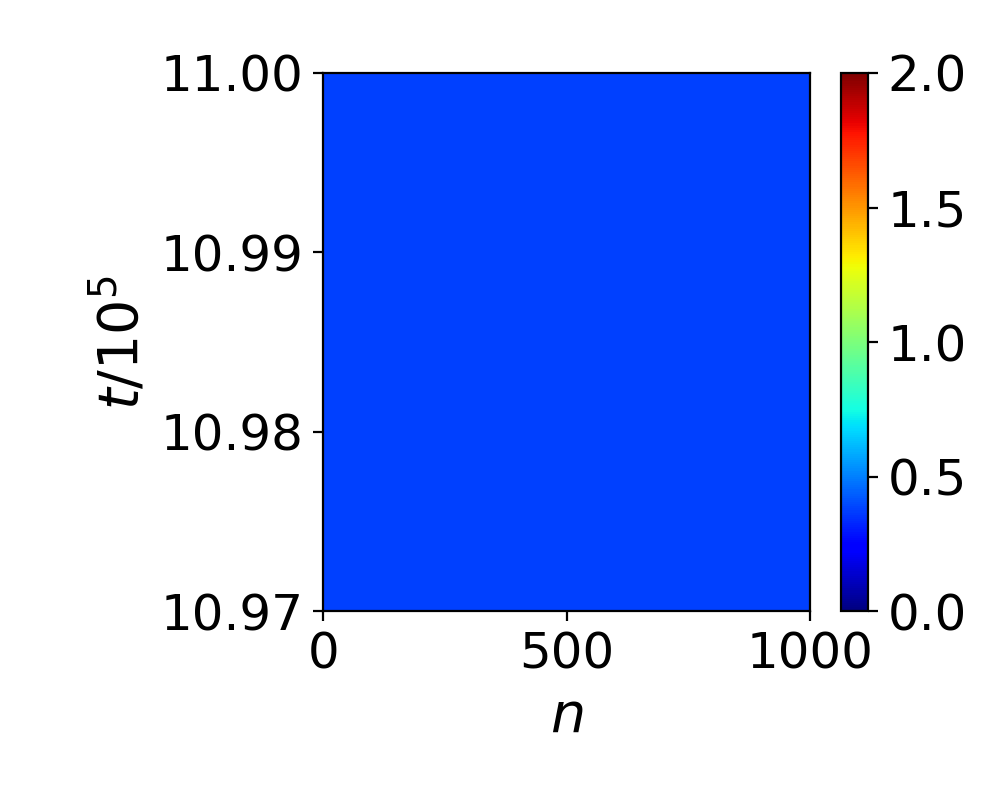}}
	\subfigure[\label{fig4:c}]{\includegraphics[width=0.51\columnwidth]{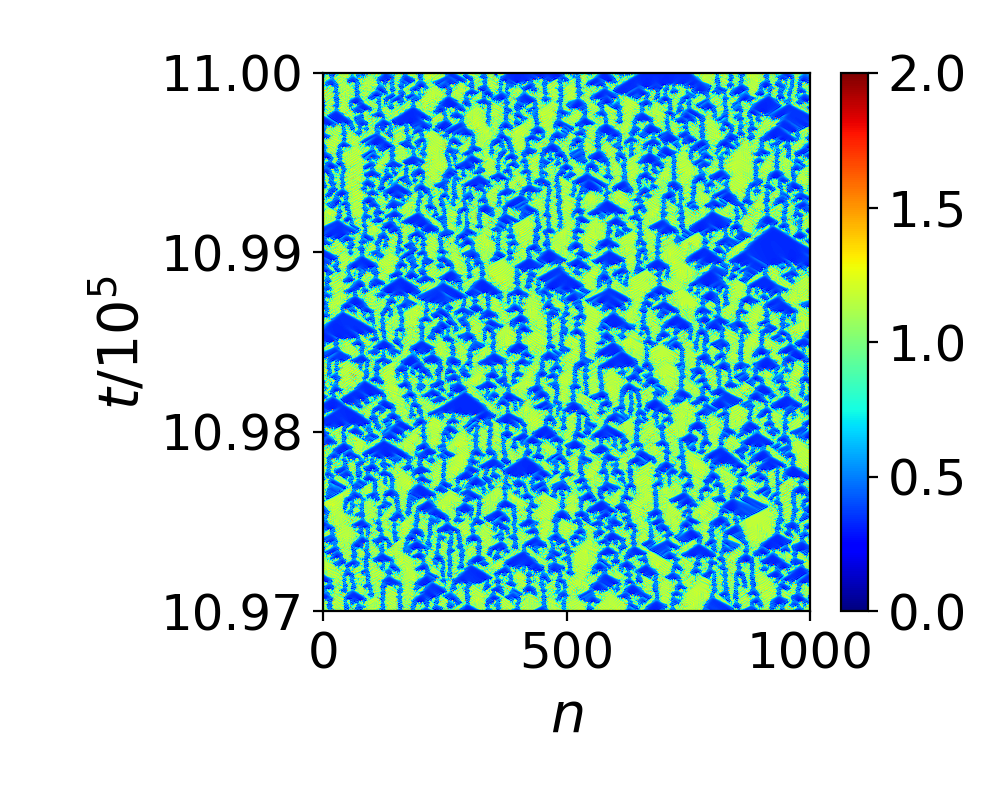}}
	\subfigure[\label{fig4:d}]{\includegraphics[width=0.51\columnwidth]{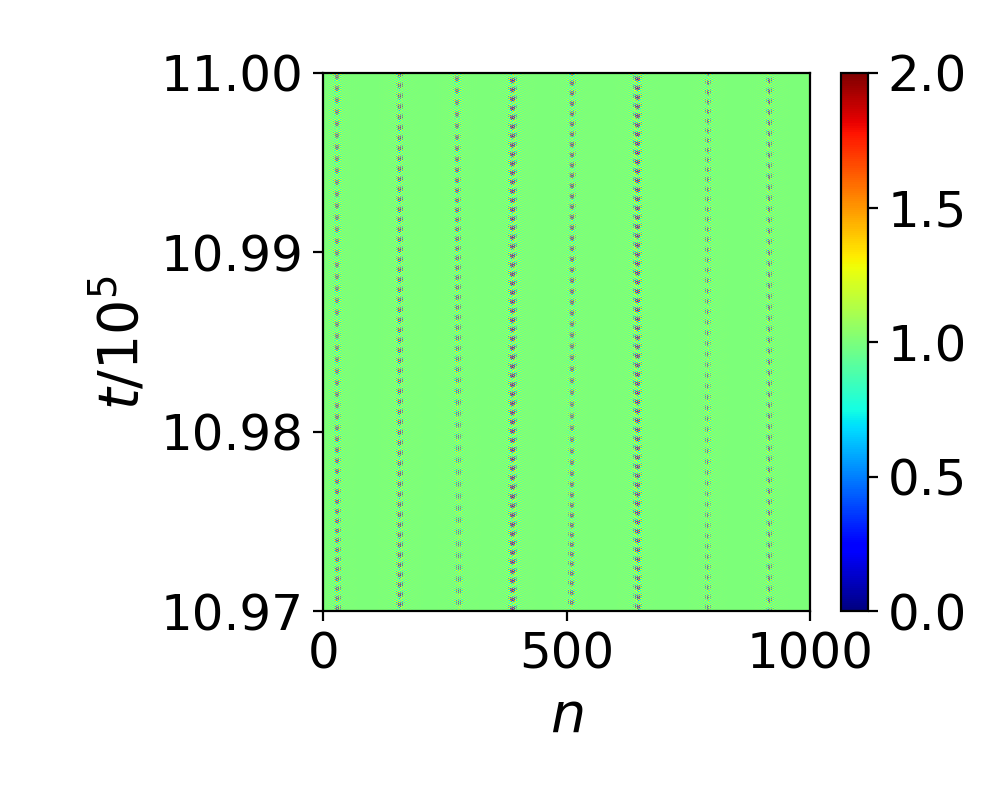}}
	\subfigure[\label{fig4:e}]{\includegraphics[width=0.51\columnwidth]{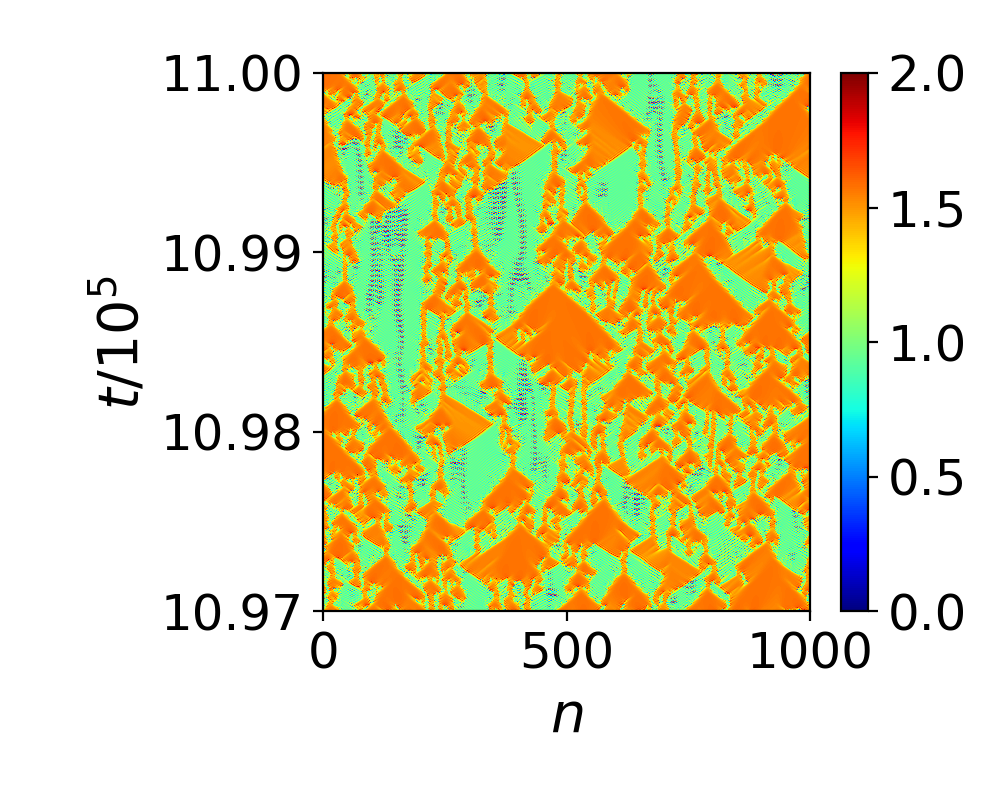}}
	\subfigure[\label{fig4:f}]{\includegraphics[width=0.51\columnwidth]{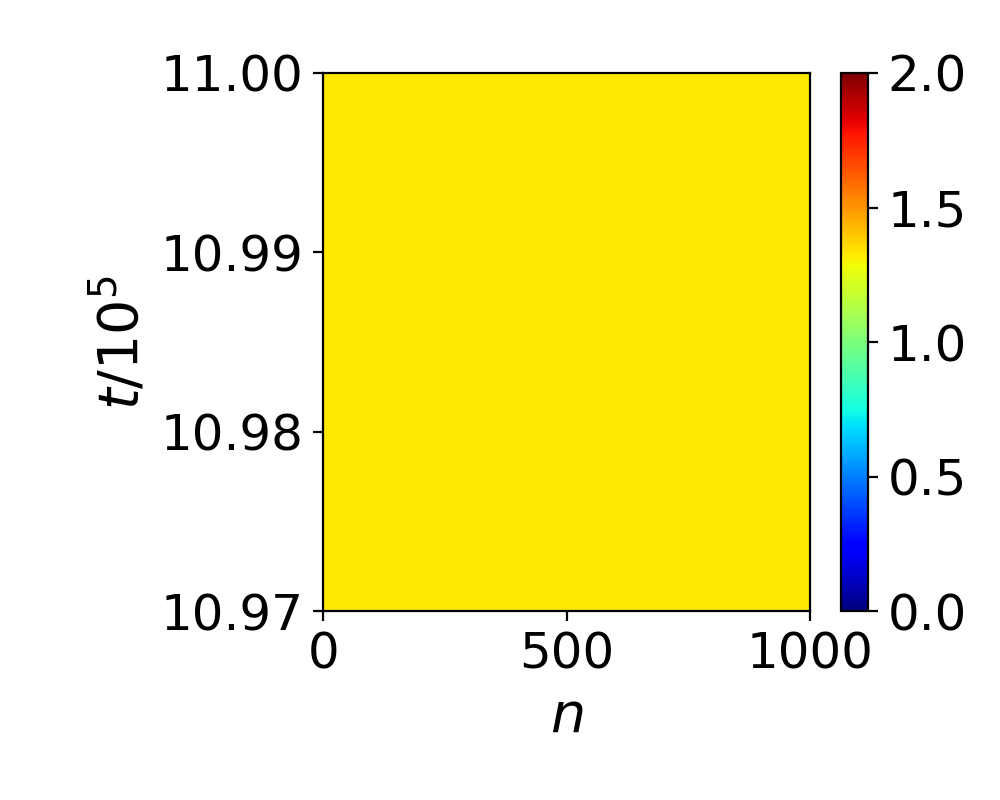}}
	\subfigure[\label{fig4:g}]{\includegraphics[width=0.51\columnwidth]{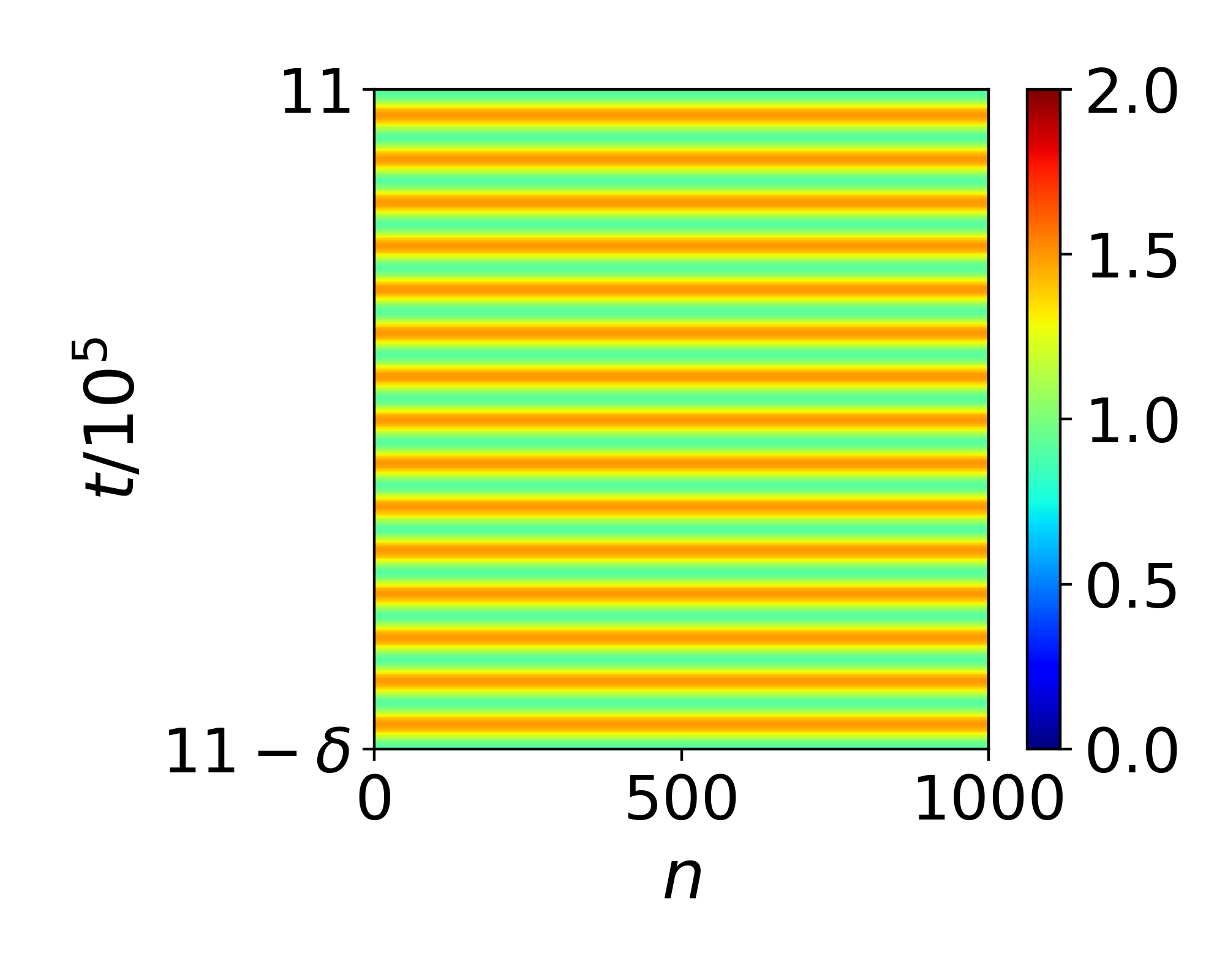}}
	\subfigure[\label{fig4:h}]{\includegraphics[width=0.51\columnwidth]{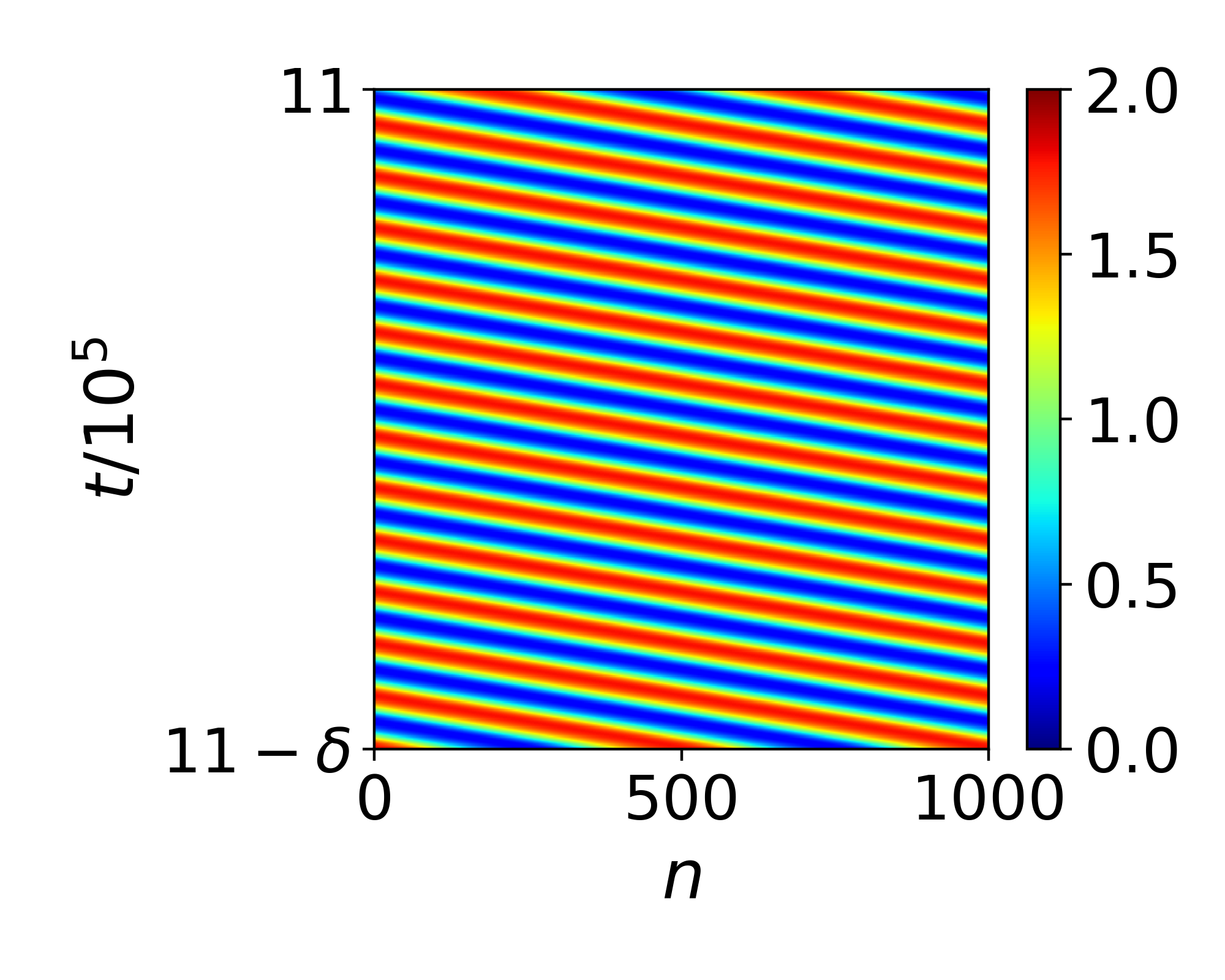}}
	\caption{Frequency evolution of the rotors with inertia value $m=1$. Each panel corresponds to different time delays: (a) $\tau=0$, (b) $\tau=1.8$, (c) $\tau=2.2$, (d) $\tau=3.1$, (e) $\tau=3.6$, (f) $\tau=4.5$, (g) $\tau=5$, and (h) $\tau=6.2$ with $\delta=0.001$ in panels (g) and (h). Other parameters as in Fig.~\ref{fig1:w-tau}.}
	\label{fig4:freqm1}
\end{figure*}
\begin{figure*}[th!]
	\centering
	\subfigure[\label{fig5:a}]{\includegraphics[width=0.51\columnwidth]{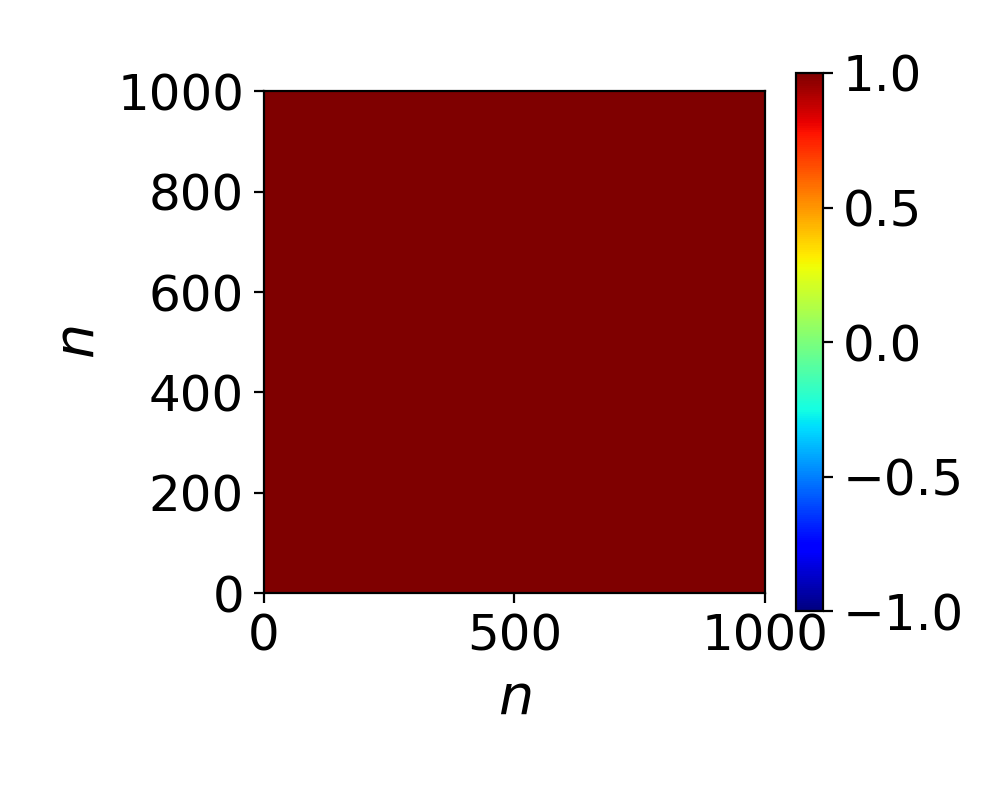}}
	\subfigure[\label{fig5:b}]{\includegraphics[width=0.51\columnwidth]{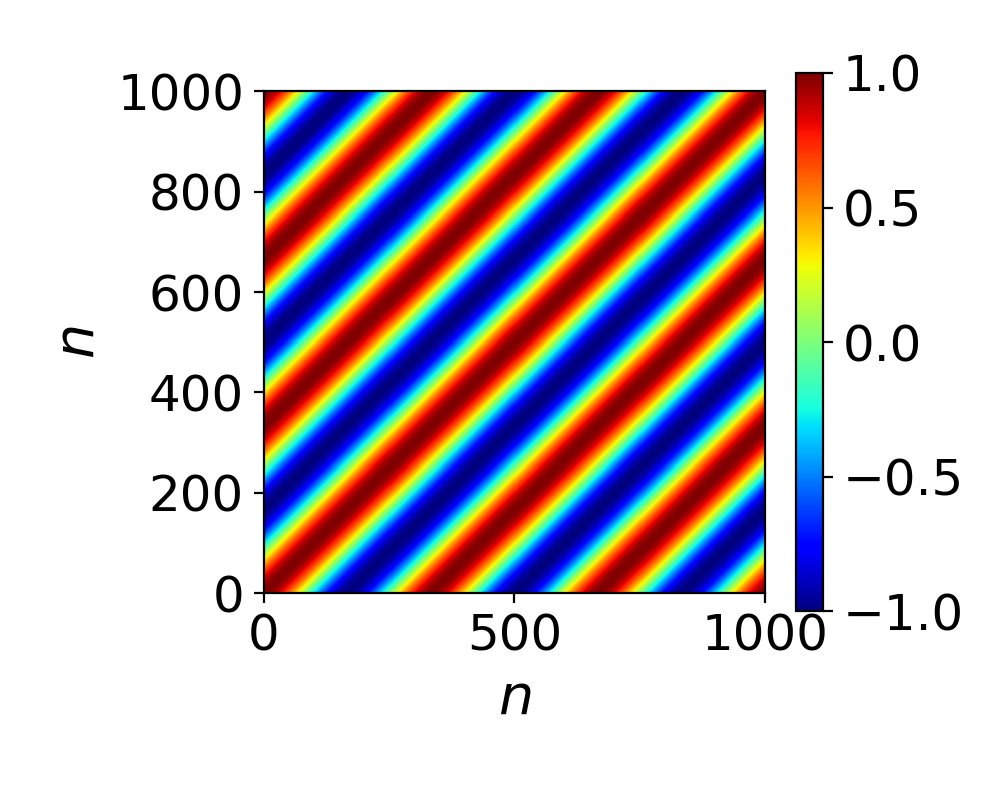}}
	\subfigure[\label{fig5:c}]{\includegraphics[width=0.51\columnwidth]{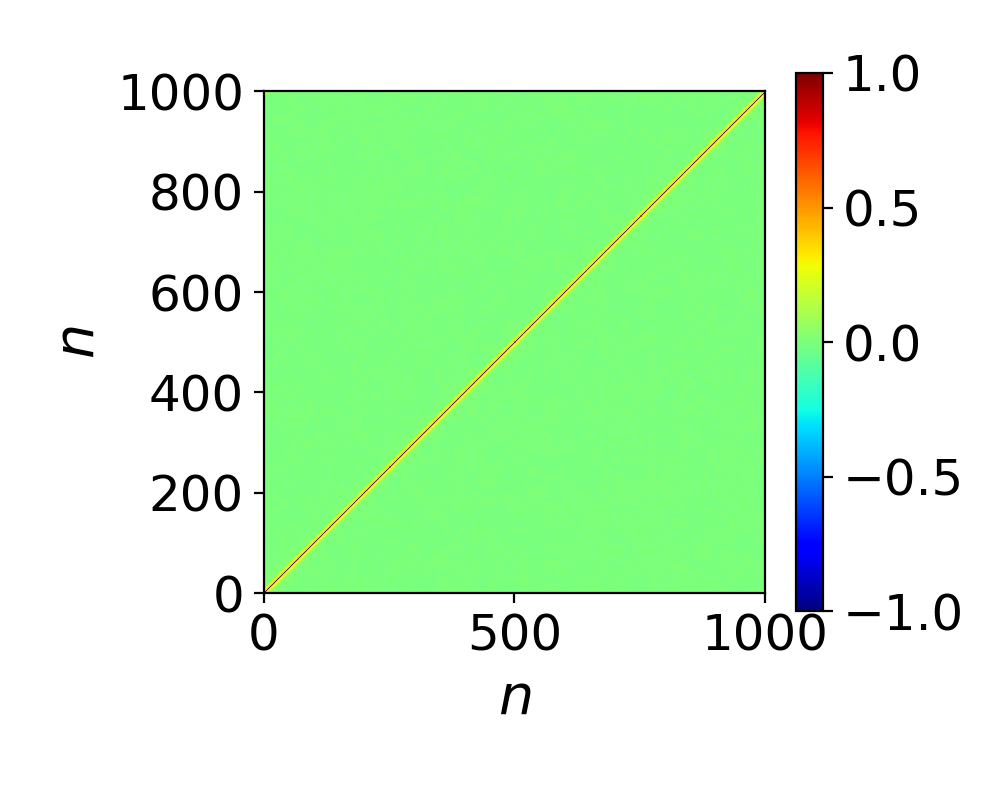}}
	\subfigure[\label{fig5:d}]{\includegraphics[width=0.51\columnwidth]{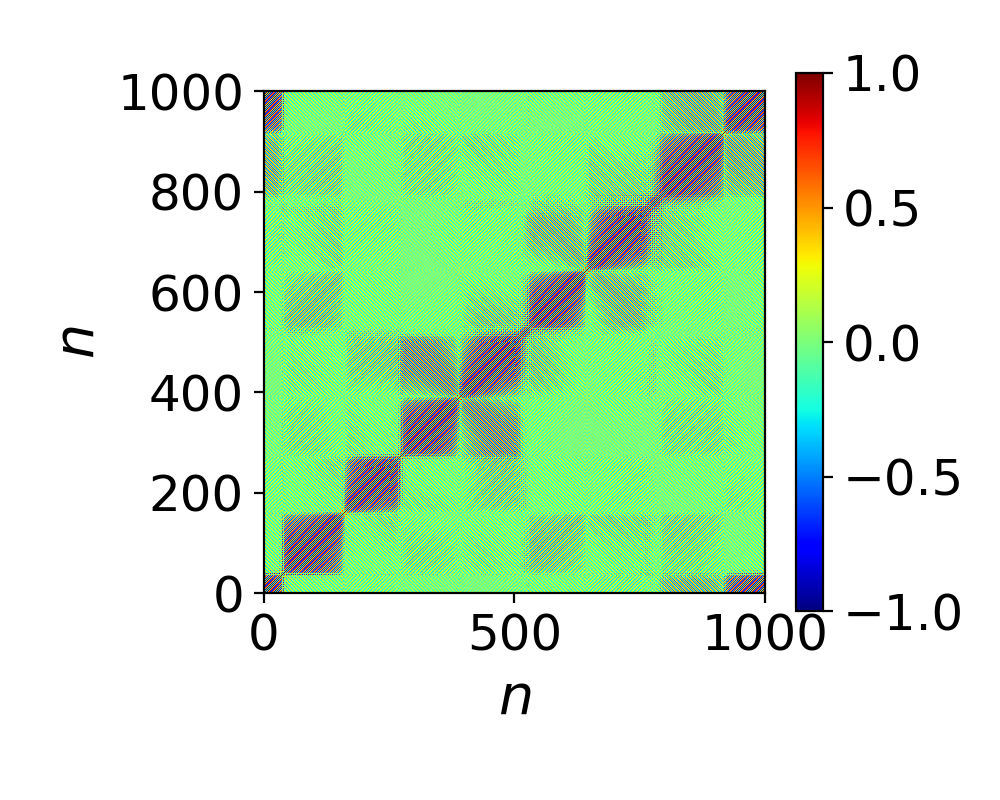}}
	\subfigure[\label{fig5:e}]{\includegraphics[width=0.51\columnwidth]{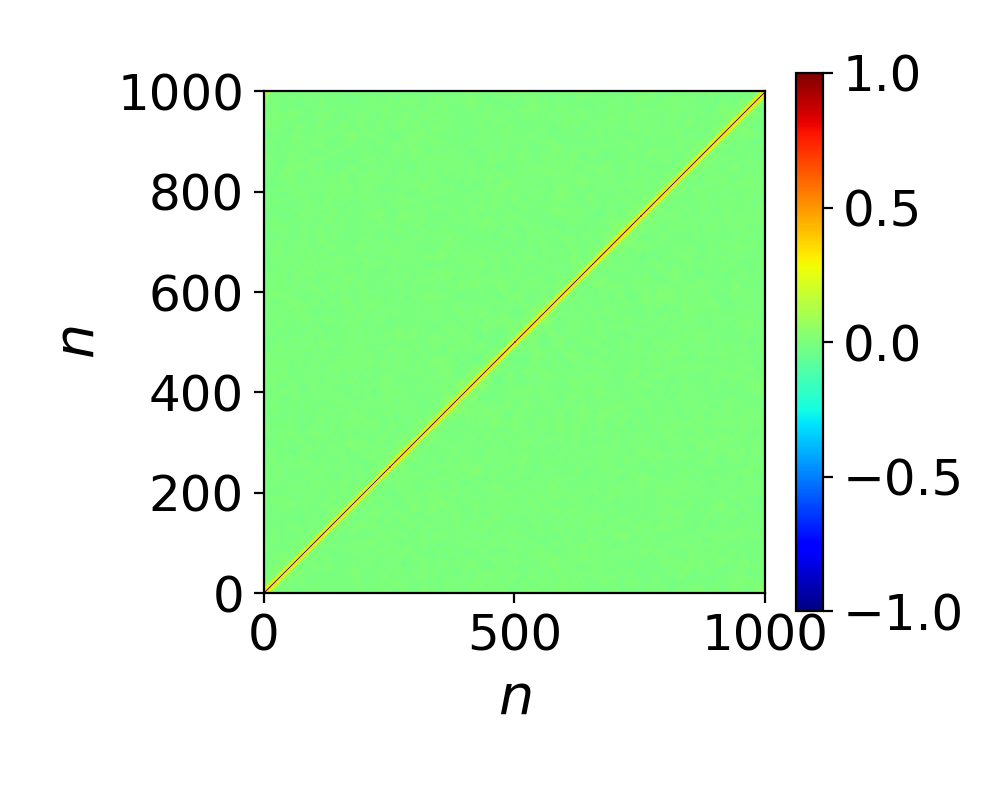}}
	\subfigure[\label{fig5:f}]{\includegraphics[width=0.51\columnwidth]{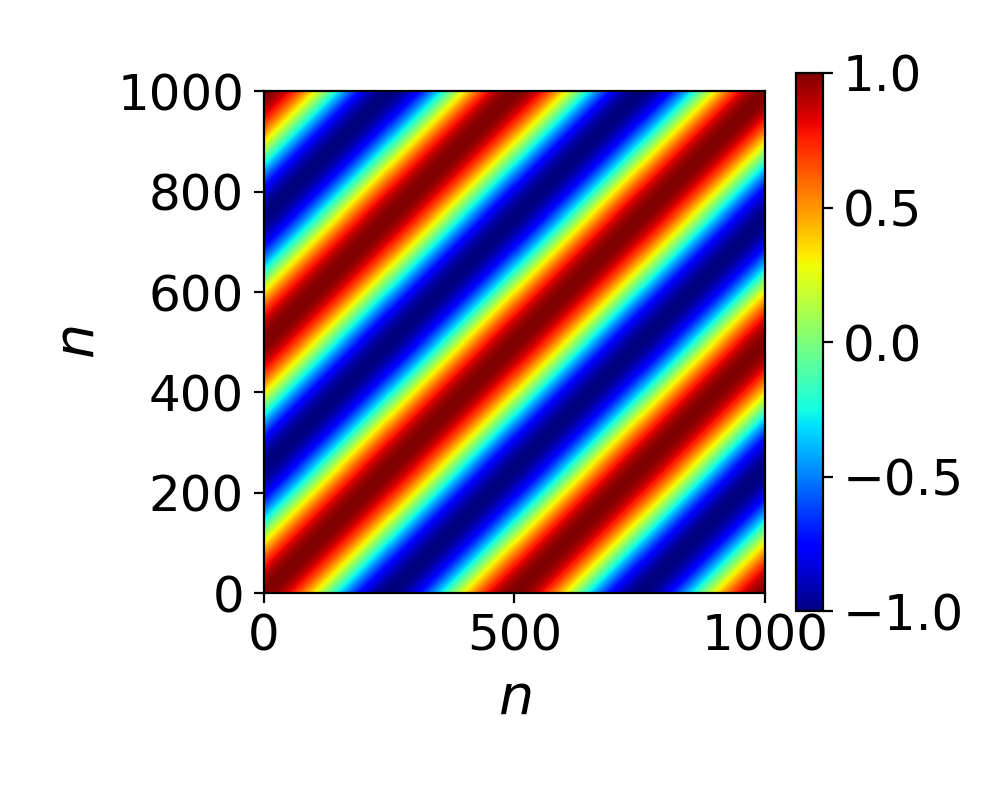}}
	\subfigure[\label{fig5:g}]{\includegraphics[width=0.51\columnwidth]{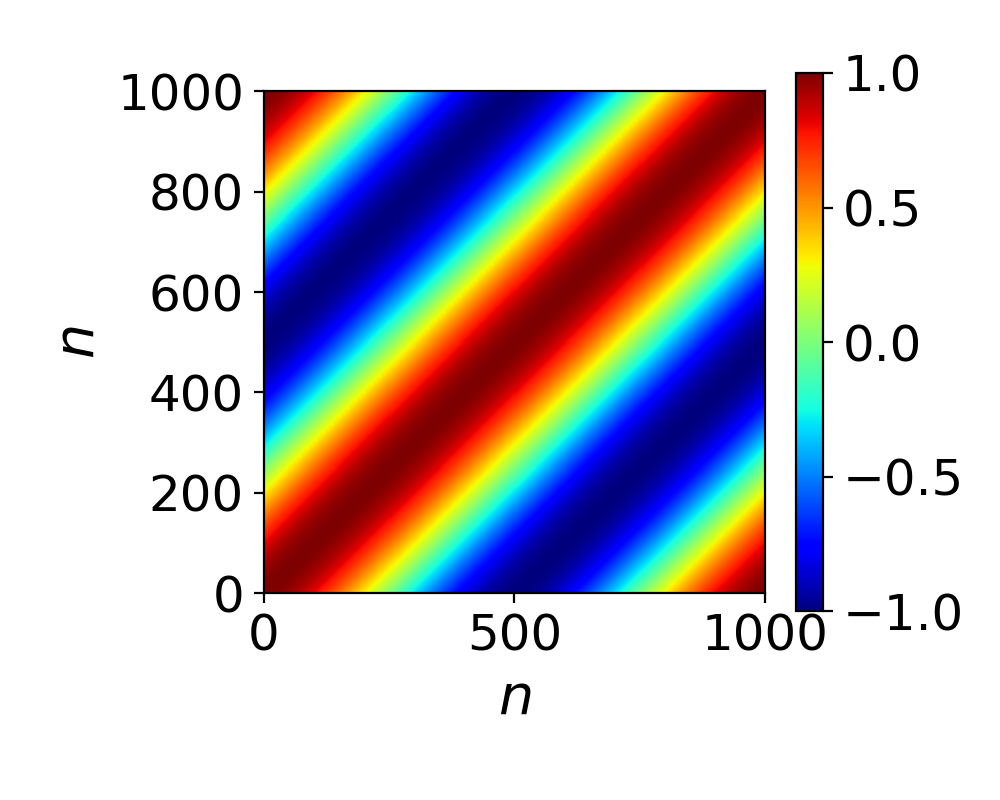}}
	\subfigure[\label{fig5:h}]{\includegraphics[width=0.51\columnwidth]{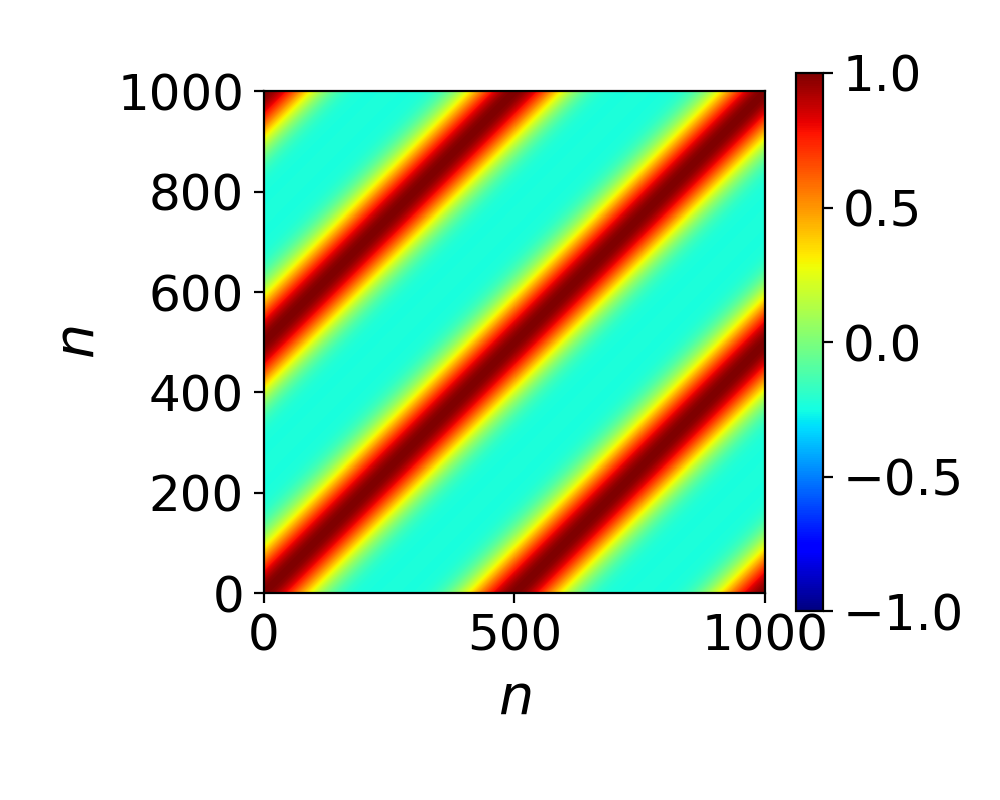}}
	\caption{Cosine similarity matrices for an inertia value $m=1$. Each panel corresponds to different time delays: (a) $\tau=0$, (b) $\tau=1.8$, (c) $\tau=2.2$, (d) $\tau=3.1$, (e) $\tau=3.6$, (f) $\tau=4.5$, (g) $\tau=5$, and (h) $\tau=6.2$. Other parameters as in Fig.~\ref{fig1:w-tau}.}
	\label{fig5:dm1}
\end{figure*}

\begin{figure*}[th!]
	\centering
	\renewcommand\thesubfigure{\fontsize{10}{10}\selectfont (\alph{subfigure})}
	\fbox{\subfigure[\label{fig7:a}]{\includegraphics[width=2.0\columnwidth]{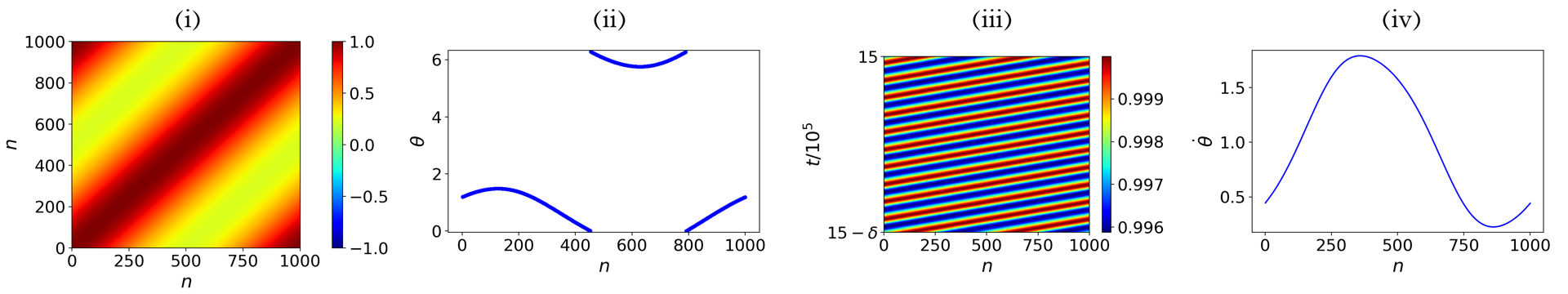}}}
	\fbox{\subfigure[\label{fig7:b}]{\includegraphics[width=2.0\columnwidth]{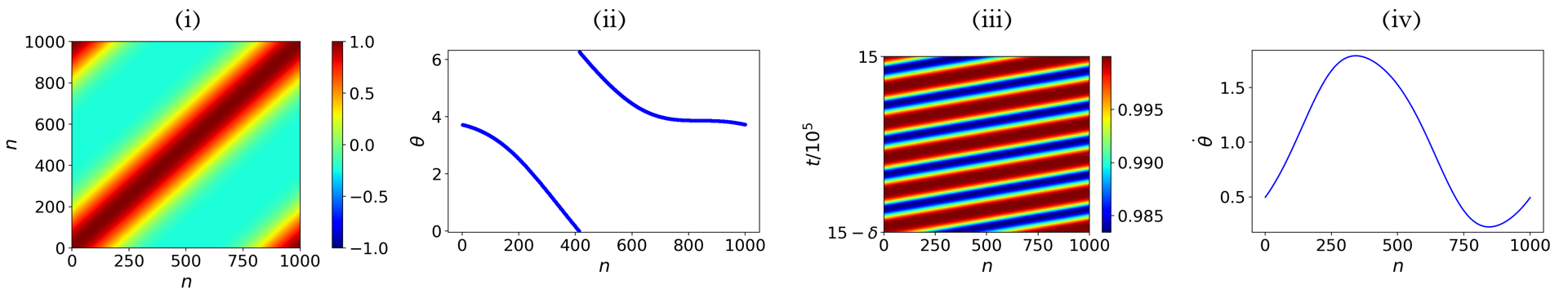}}}
	\fbox{\subfigure[\label{fig7:c}]{\includegraphics[width=2.0\columnwidth]{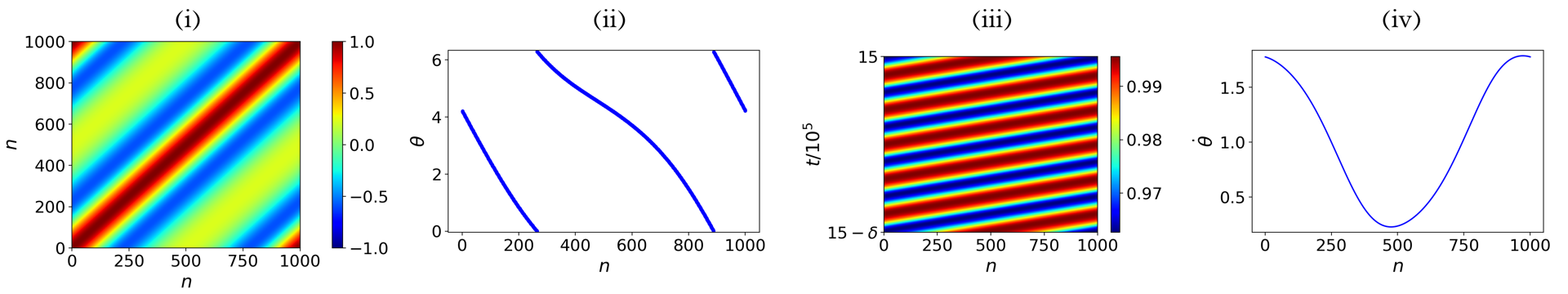}}}
	\fbox{\subfigure[\label{fig7:d}]{\includegraphics[width=2.0\columnwidth]{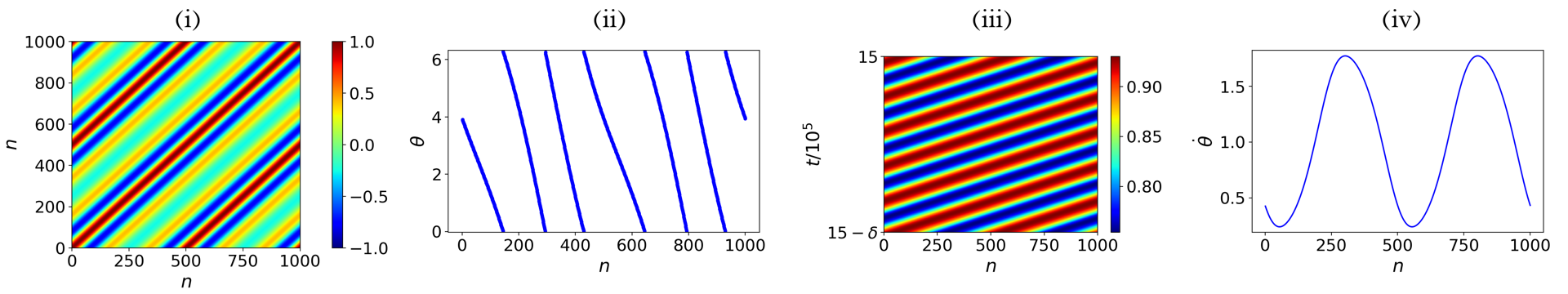}}}	
	\caption{(i) Cosine similarity matrices, (ii) phase snapshots, (iii) evolution of the local order parameter ($\delta=0.0004$), and (iv) snapshots of angular velocities for an with inertia value $m=1$, and time delay $\tau=6.2$. The order parameters correspond to (a) $\langle r \rangle_\infty=0.766$, (b) $\langle r \rangle_\infty=0.438$, (c) $\langle r \rangle_\infty=0.117$, and (d) $\langle r \rangle_\infty=0.017$. Other parameters as in Fig.~\ref{fig1:w-tau}.}
	\label{fig7:m1}
\end{figure*}

\begin{figure*}[th!]
	\centering
	\subfigure[\label{fig6:a}]{\includegraphics[width=0.51\columnwidth]{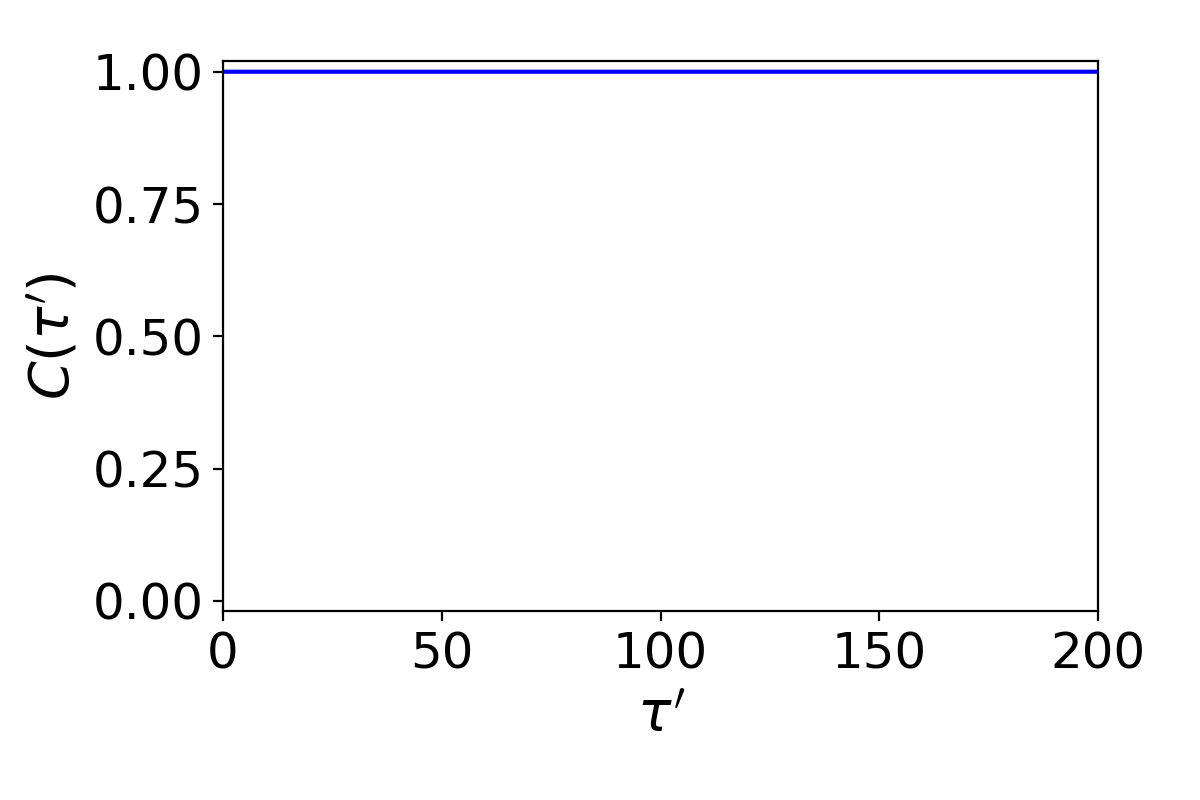}}
	\subfigure[\label{fig6:b}]{\includegraphics[width=0.51\columnwidth]{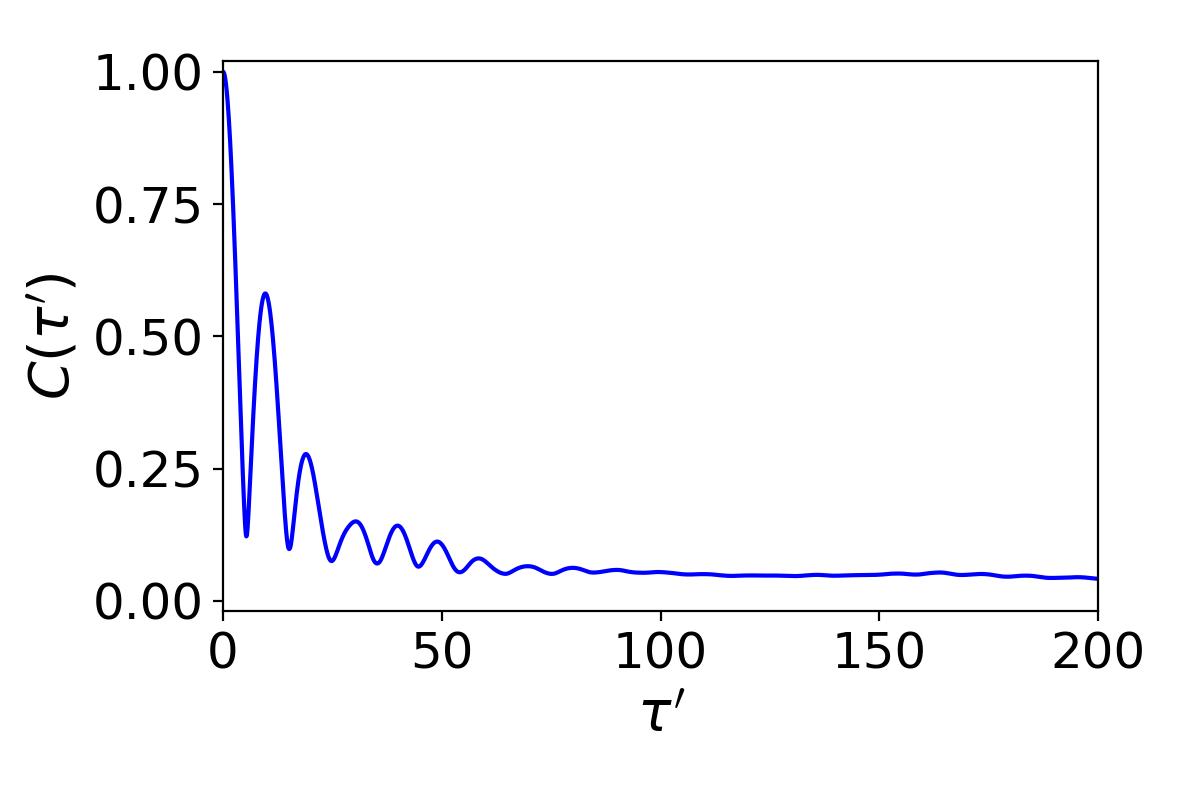}}
	\subfigure[\label{fig6:c}]{\includegraphics[width=0.51\columnwidth]{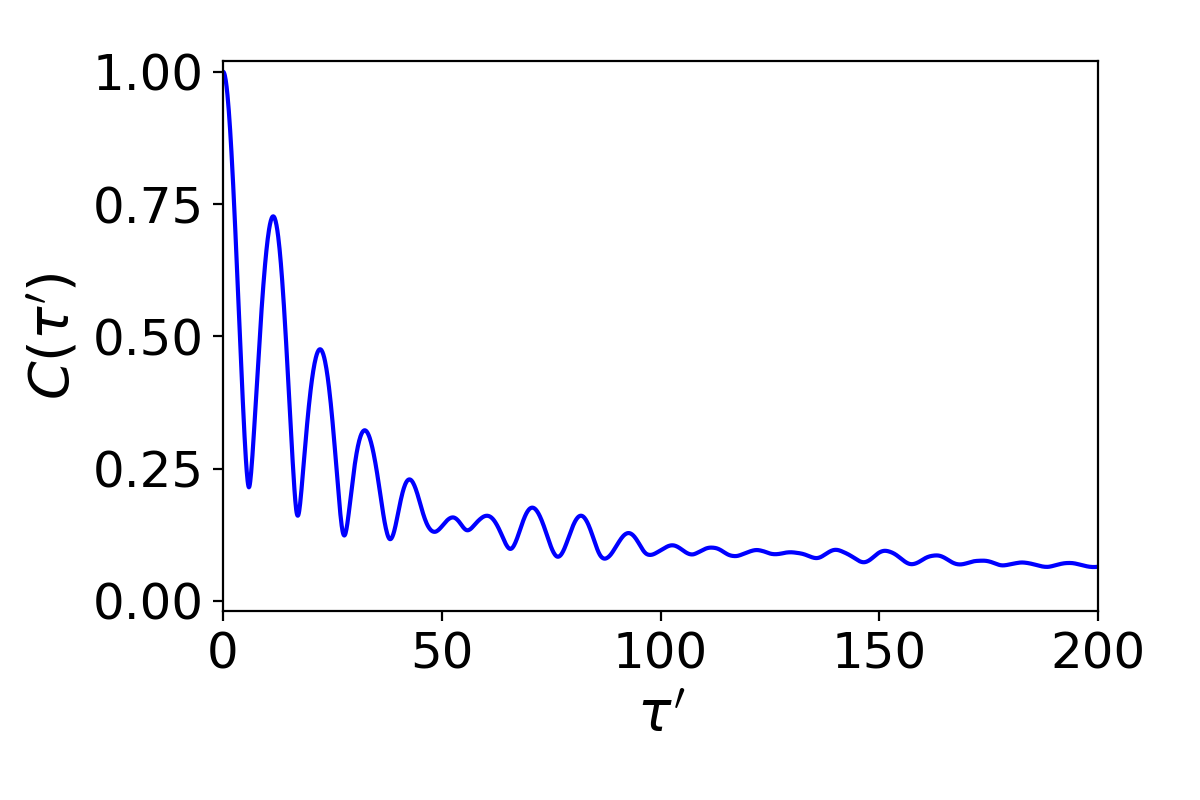}}
	\subfigure[\label{fig6:d}]{\includegraphics[width=0.51\columnwidth]{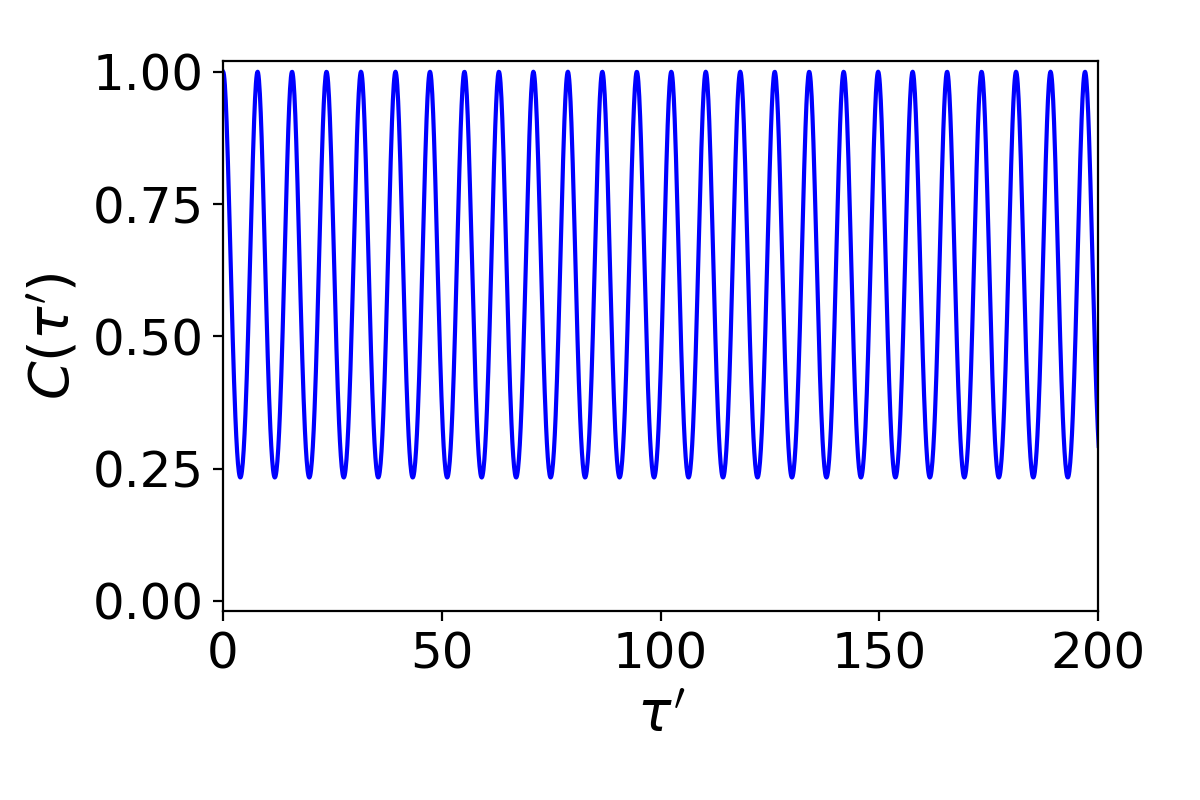}}
	\caption{Autocorrelation function with inertia value $m=1$ for the same initial conditions. Each panel corresponds to: (a) $\tau=1$, (b) $\tau=2.2$, (c) $\tau=3.6$, and (d) $\tau=6.2$. In panel (d), the averaged order parameter $\langle r \rangle_\infty$ equals $0.438$. Other parameters as in Fig.~\ref{fig1:w-tau}.}
	\label{fig6:outo}
\end{figure*}

\begin{figure}[]
	\centering
	\subfigure[\label{fig8:a}]{\includegraphics[width=1\columnwidth]{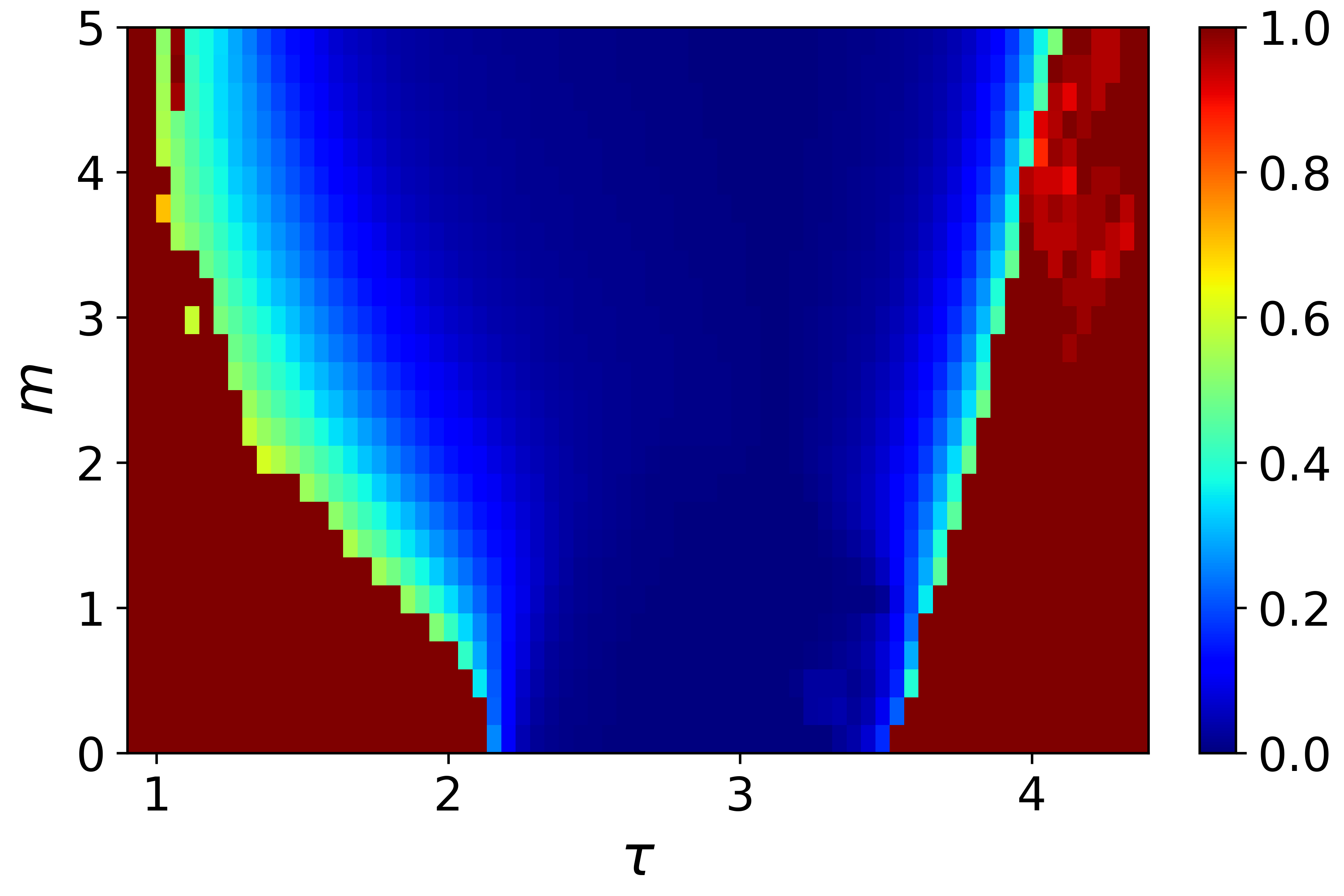}}
	\subfigure[\label{fig8:b}]{\includegraphics[width=1\columnwidth]{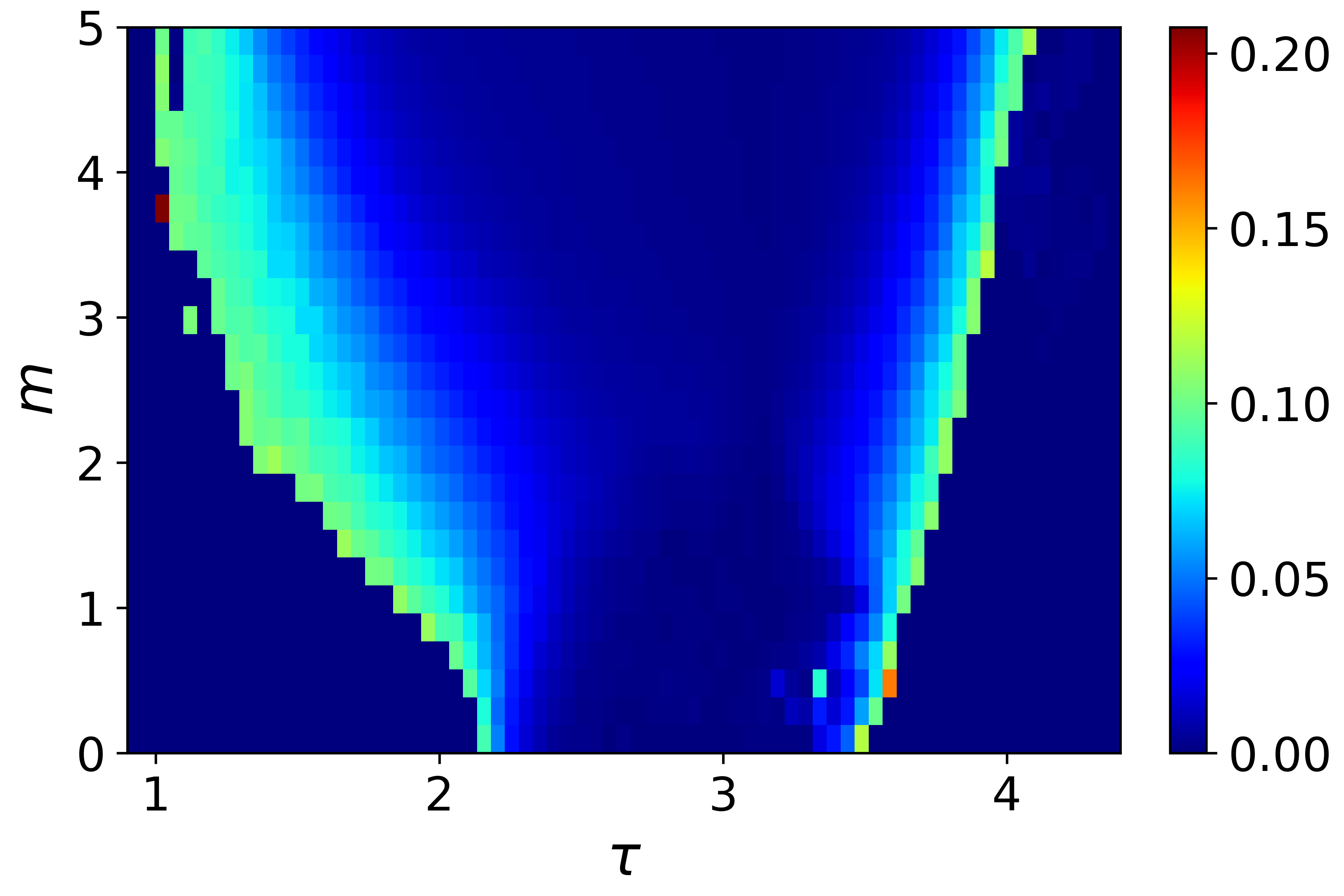}}
	\caption{(a) The time-averaged spatial correlation function $g_0(t)$ and (b) the standard deviation of its fluctuations versus time delay and inertia. Other parameters as in Fig.~\ref{fig1:w-tau}.}
	\label{fig8:heatmapchimera}
\end{figure}

\begin{figure*}[th!]
	\centering
	\renewcommand\thesubfigure{\fontsize{10}{10}\selectfont (\alph{subfigure})}
	\fbox{\subfigure[\label{fig10:a}]{\includegraphics[width=1\textwidth]{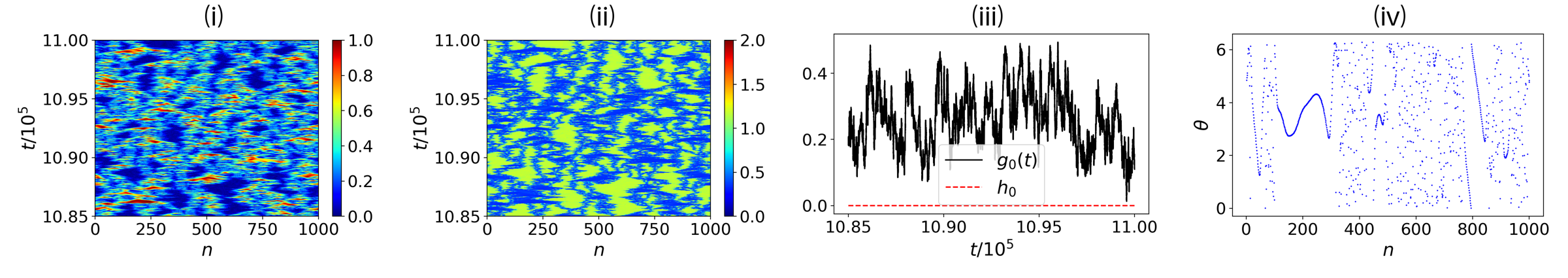}}}
	\fbox{\subfigure[\label{fig10:b}]{\includegraphics[width=1\textwidth]{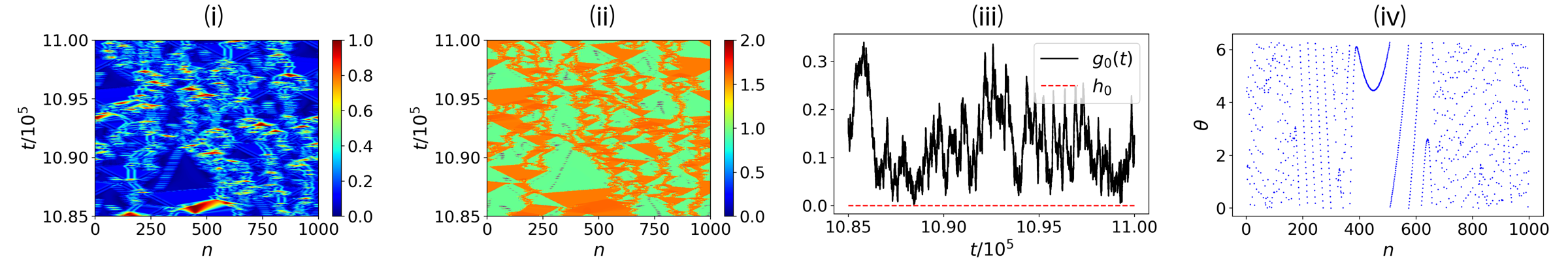}}}
	\fbox{\subfigure[\label{fig10:c}]{\includegraphics[width=1\textwidth]{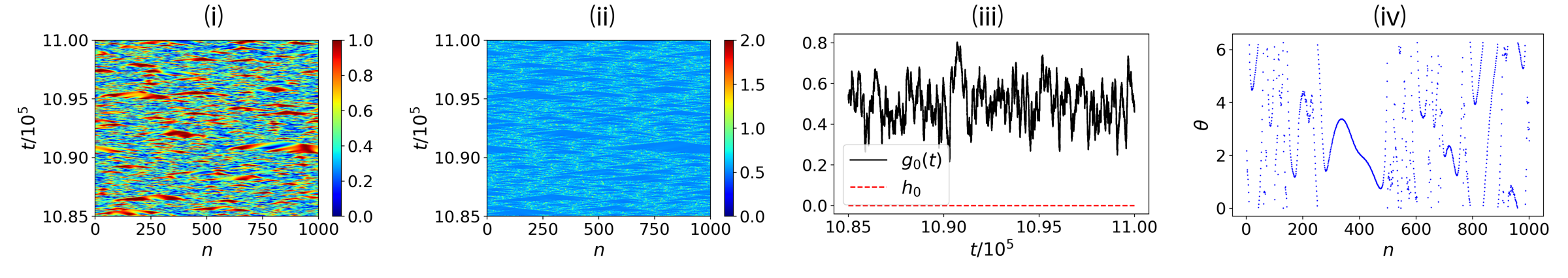}}}
	\fbox{\subfigure[\label{fig10:d}]{\includegraphics[width=1\textwidth]{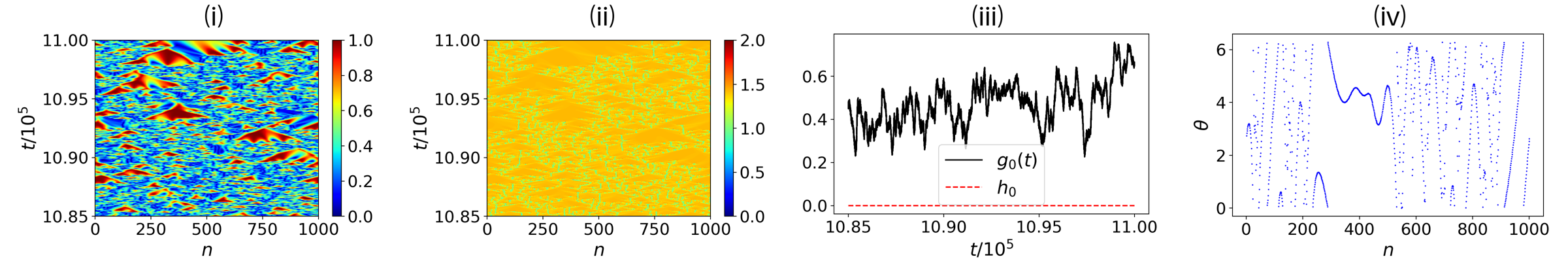}}}
	 \caption{Time evolution of (i) the local order parameter, (ii) angular velocities, (iii) $g_0(t)$ values along with the value of $h_0$, and (iv) snapshot of phases. Each panel corresponds to a different set of parameters: (a) $m=0$, $\tau=2.15$; (b) $m=0$, $\tau=3.5$; (c) $m=5$, $\tau=1$; and (d) $m=5$, $\tau=4.1$. Other parameters as in Fig.~\ref{fig1:w-tau}.}
	\label{fig10:chimera}
\end{figure*}

To examine the effects of inertia and time delay on synchronization dynamics, we performed analytical and numerical studies on a regular network of $N=1000$ rotors with an average degree $\langle k \rangle=10$. For simplicity, the coupling strength ${\alpha}$ and intrinsic frequency $\omega$ were set to unity. The results of these investigations are presented below.

Our analysis of phase-locked solutions in fully synchronized states highlights key aspects of the system dynamics:
While the number of possible phase-locked solutions is determined exclusively by network topology and remains invariant under changes in inertia, the stability of these solutions exhibits strong dependence on both inertia $m$ and time delay $\tau$. Figure~\ref{fig1:w-tau}~(a) illustrates this behavior through the evolution of stable phase-locked frequencies $\Omega_f$ across different inertia values ($m = 0, 0.6, 1, 5, 10$). Three characteristic regimes emerge: (i) multistability arises when $\tau \gtrsim 2.4$, allowing multiple stable solutions to coexist; (ii) finite inertia systematically destabilizes solutions compared to the inertia-free case, with the effect intensifying at larger delays; and (iii) the inertia-dependent stability exhibits a non-monotonic pattern, reflecting complex nonlinear interactions that selectively shrink the basins of synchronization in a parameter-dependent manner.

We numerically verified these analytical predictions through extensive simulations using a sixth-order predictor-corrector integration scheme with a fixed time step of $dt = 0.01$. To ensure the system reached a statistically stationary state, simulations are run for $10^8$ time steps, followed by averaging over $10^7$ steps. Since the basins of attraction for many phase-locked solutions are extremely small, numerous solutions cannot be accessed through random initialization~\cite{ameli2025synchronization,mahdavi2025synchronization}. Therefore, the initial phases, system history, and angular velocities were carefully chosen to capture the first solution from each frequency branch shown in Fig.~\ref{fig1:w-tau}~(a). Figure~\ref{fig1:w-tau}~(b) displays numerical results obtained via forward parameter continuation, in which each solution branch was tracked using the endpoint of the previous computation as the subsequent initial condition. This approach not only confirms the analytically predicted topology-governed multiplicity of solutions but also validates the complex inertia-induced stability modifications, as evidenced by the excellent agreement between theory and simulation across the full parameter range. The representation is restricted to fully synchronous linear phase-locked solutions.

Next, we extend our numerical analysis to investigate the effects of time delay in systems with inertia but without memory effects. The initial phases $\theta_i(0)$ are randomly sampled from a uniform distribution over $[-\pi, \pi]$, while the initial angular velocities $\dot{\theta}_i(0)$ are set to zero. To account for time-delayed interactions, the simulation is configured so that rotors evolve independently (without coupling) during the initial interval $[0, \tau]$. For times $t > \tau$, the interactions at time $t$ depend on the rotor states at the earlier time $t-\tau$, thereby naturally incorporating the system’s history into its dynamics.

Figure~\ref{fig2:r-tau0} provides a systematic numerical characterization of the delay-dependent stationary states across different inertia regimes. The panel layout highlights parameter-dependent transitions, with each row corresponding to a specific inertia value and columns representing four key aspects of stationary collective dynamics: (i) the stationary Kuramoto order parameter $\langle r \rangle_\infty$, which quantifies the degree of global phase synchronization in the long-time limit; (ii) $\mathrm{\sigma}_r$, capturing fluctuations of the order parameter in the stationary regime; (iii) the stationary ensemble-averaged mean rotor frequency $\langle \dot{\theta} \rangle_\infty$; and (iv) $\mathrm{\sigma}_{\dot{\theta}}$, measuring the dispersion of individual rotor frequencies in the asymptotic state.

To ensure statistical robustness, we performed 60 independent realizations for each parameter set, each with distinct random initial conditions. The results show both individual trajectories (red dots) and ensemble statistics (blue dots with error bars corresponding to $\pm\sigma/\sqrt{60}$ standard error). This dual representation captures both the distribution of possible stationary states across different initial conditions and their mean behavior with quantified confidence. By combining multiple metrics across well-sampled realizations, this analysis allows a comprehensive identification of synchronization transitions and stability thresholds as a function of $\tau$.

The first column shows the steady-state order parameter as a function of time delay for different inertia values. An intermediate time-delay regime emerges in which synchronization is significantly weakened, approaching an effectively asynchronous state. Within this regime, the order parameter is largely insensitive to ensemble variations. Notably, the width of this desynchronized region increases systematically with inertia.
At the left and the right of this intermediate region, the system exhibits bistability, alternating between full synchronization and complete desynchronization, with the final state strongly dependent on initial conditions. This multistability underscores the system’s sensitivity to its initial preparation.

The second column quantifies the temporal evolution of the order parameter via its standard deviation over time, averaged across ensembles. In phase-locked regimes -- both synchronized and desynchronized -- the order parameter remains constant, yielding a zero standard deviation, in excellent agreement with the analytical predictions in Fig.~\ref{fig1:w-tau}~(a). In contrast, within the intermediate time-delay region, the order parameter shows pronounced temporal fluctuations, which are largest near the region boundaries and increase markedly with inertia.

In the third column, we show the ensemble-averaged mean oscillator frequency in the steady state. Among the synchronous solution branches identified analytically (Fig.~\ref{fig1:w-tau}~(a)), phase-locked states near the intrinsic frequency dominate numerically, reflecting their larger basins of attraction. However, within the intermediate parameter region, which broadens with increasing inertia, no phase-locked solutions are observed across the 60 ensemble realizations. This absence arises because the basins of attraction for these synchronous states are extremely narrow, requiring precise initial conditions. As a result, increasing inertia expands the regime dominated by non-phase-locked dynamics. At the boundaries of this intermediate region, inertia-free systems exhibit abrupt jumps in mean frequency between solutions. With increasing inertia, these transitions smooth out, suggesting that inertia regularizes frequency dynamics. Notably, the frequencies corresponding to the fully synchronized ($\langle r \rangle_\infty=1$) and fully asynchronous ($\langle r \rangle_\infty=0$) states are very close, supporting the proposition that linear phase-locked solutions in both states can attain similar values, as demonstrated in the supplementary mathematical proof and illustrated in Figs.~{SF1} and {SF2}.

To further examine individual oscillator behavior, we analyze the statistical distribution of rotor frequencies over steady-state trajectories. The fourth column presents the standard deviation of this distribution. In phase-locked states, the dispersion vanishes, indicating complete frequency entrainment. Within the intermediate region, finite standard deviations reflect inter-oscillator frequency differences, temporal variations, or both. Frequency dispersion peaks near the region boundaries, consistent with the previously observed maximal order parameter fluctuations. Importantly, substantial frequency heterogeneity emerges among oscillators within this region, particularly near its edges.

For $m=1$, a particularly notable dynamical regime emerges when the time delay exceeds $\tau \approx 4.85$. In this range, we observe solutions covering the full interval $r \in [0,1]$ that appear uniquely for this inertia value. Multiple stable solutions coexist, each maintaining a time-invariant order parameter despite ongoing variations in individual rotor frequencies (see Fig.~\ref{fig2:r-tau0}~(c)). This transition coincides with the destabilization of the second frequency branch of phase-locked solutions shown in Fig.~\ref {fig1:w-tau}~(a)~(iii), which previously dominated the dynamics due to their large basins of attraction near the intrinsic frequency. The loss of stability of these solutions allows the emergence of novel states not captured by linear approximations. This phenomenon will be examined in greater detail later.

Since the inertial state $m=1$ displays richer and more complex dynamics than other inertia values, we focused our analysis on this regime to examine the local behavior of network oscillators. Specifically, we computed the distribution of angular velocities, the temporal evolution of these velocities, and the cosine similarity matrices under varying time delays. The corresponding results are shown in Figs.~\ref{fig3:pwm1}--\ref{fig5:dm1}.

In Fig.~\ref{fig3:pwm1}, the distribution functions of angular velocities in panels (a), (b), and (f) are obtained in the form of the Dirac delta function, indicating that the angular velocity of all oscillators remains constant in time and equal to each other. This case demonstrates the existence of a linear phase-locked solution in the network, and the time evolution of the corresponding angular velocities in Fig.~\ref{fig4:freqm1} confirms this result. This also demonstrates that the oscillator's angular velocity depends on the time delay. Additionally, the corresponding cosine similarity matrices in Fig.~\ref{fig5:dm1} reveal that these linear phase-locked solutions converge either to $\langle r \rangle_\infty = 1$ (panel a) or to $\langle r \rangle_\infty = 0$ (panels b and f), with the latter corresponding to helical phase patterns.

Panels (c), and (e) of Fig.~\ref{fig3:pwm1} exhibit two-peaked distribution functions, whose corresponding time evolution in Fig.~\ref{fig4:freqm1} indicates the variation of angular velocities of the oscillators between the two peaks. This behavior, where each rotor evolves independently and separately from the others, results in the vanishing similarity between them, as seen in Fig.~\ref{fig5:dm1}.

The results in Fig.~\ref{fig3:pwm1}~(d) and also show unimodal angular frequency distribution functions with finite width.
As shown by the time evolution of the corresponding angular velocities in Fig.~\ref{fig4:freqm1}, some oscillators evolve with identical angular velocities while others change their angular velocities over time. This behavior leads to the formation of two groups of correlated and uncorrelated oscillators, as illustrated in the corresponding similarity matrices in Fig.~\ref{fig5:dm1}.

Finally, Figs.~\ref{fig3:pwm1}~(g) and ~\ref{fig3:pwm1}~(h) display $U$-shaped distribution functions of angular velocities for $\tau=$ 5 and 6.2, respectively. This frequency distribution has not been previously observed in regular networks of first-order Kuramoto oscillators~\cite{ameli2025synchronization}. The $U$-shaped distribution of angular velocities indicates the periodic variation of  the angular frequency of oscillators over time. 
It has to be noted that the nature of the angular velocity evolution depends on the value of the order parameter. When $\langle r \rangle_\infty$ equals zero or one, all oscillators evolve identically (Fig~\ref{fig4:freqm1}~(g) and see the supplementary video SV1). 
However, when $\langle r \rangle_\infty$ has a value between zero and one (as seen in Figs.~\ref{fig4:freqm1}~(h) and ~\ref{fig5:dm1}~(h)), the rotors no longer have equal frequencies at a given time, and the angular velocity evolution diagram turns to tilted strips (Fig~\ref{fig4:freqm1}~(h)). This state is a travelling wave in which some rotors get closer together and others move apart, resulting in a partially synchronized state with $0<\langle r \rangle_\infty<1$ (see the supplementary video SV2 for a clear illustration of the dynamics). Consequently, the similarity matrix in this case closely resembles the helical patterns with $\langle r \rangle_\infty=0$. However, due to temporal changes in the phase difference of adjacent rotors, their elements undergo modifications, as visible in Fig.~\ref{fig5:dm1}~(h). 

Figure~\ref{fig7:m1} elaborates further on the case of $\tau = 6.2$ and  $m = 1$ and displays similarity matrices, phase snapshots, time evolutions of local order parameter, and angular velocity snapshots corresponding to four distinct behaviors with $\langle r \rangle_\infty = 0.766, 0.438, 0.117$, and $0.017$.
Panels (a)--(c) exhibit a single-cycle behavior, each with one curved phase line, whereas panel (d) shows two cycles, producing two curved lines and larger-scale periodicity.
All panels exhibit traveling wave behavior, in which rotors periodically converge and diverge while the global order parameter remains constant. This is reflected in the diagonal-line patterns of the local order parameter.


Figure~\ref{fig6:outo} presents the autocorrelation function for $m = 1$ and various time delays. Panel (a) represents the fully coherent states, that is, the phase-locked states with $\langle r \rangle_\infty = 0$ or $1$. Panels (b) and (c) showing short-range correlation represent the chaotic states emerging at the transition points. Finally, panel (d) represents a travelling wave state with $0<\langle r \rangle_\infty <1$ whose periodicity arises from the rotors' regularly oscillating angular velocities.

Next, we investigate the effects of time delay and inertia on chimera state formation. Using the methodology of Kemeth et al.~\cite{kemeth2016classification}, we employ the spatial correlation function $g_0(t)$ for quantitative characterization. Both the time-averaged value and temporal standard deviation of $g_0(t)$ are computed to provide a comprehensive assessment of the system’s behavior.

Figure~\ref{fig8:heatmapchimera} presents the mean and standard deviation of $g_0(t)$ across various inertia and time delay values. Regions with $\langle g_0(t)\rangle = 1$ indicate non-chimera states, corresponding to either complete synchronization or the helical states. In synchronized states, all rotors share identical phases, whereas the helical states ($\langle r \rangle_\infty = 0$) show phases uniformly distributed around a trigonometric circle with minimal separation. In both cases, we find $g_0(t) = 1$ with zero standard deviation, indicating stable spatial correlations and the absence of chimera states.

For higher inertia and time-delay values, $\langle g_0(t)\rangle$ approaches unity with minor deviations, indicating phase-locked states in which most rotors maintain regular spacing while a few evolve independently. The near-zero standard deviation of $g_0(t)$ in these regions further confirms the absence of significant chimera states, as the system primarily exhibits modified phase-locked behavior rather than true chimera patterns.


The remaining regions where the spatial correlation function satisfies $0 < \langle g_0(t)\rangle < 1$ correspond to chimera states in the network, characterized by the coexistence of coherent and incoherent domains. Within these regions, the magnitude of $\langle g_0(t)\rangle$ directly reflects the spatial extent of coherence, with larger values indicating more extensive synchronized domains and consequently larger chimera states. Fig.~\ref{fig8:heatmapchimera}~(a) reveals two distinct parameter regions supporting chimera state formation, both of which expand with increasing inertia $m$. Notably, inertia modifies the characteristic time delays for chimera emergence: at the left region shifts toward shorter delays while at the right zone moves towards higher delays as $m$ increases. This demonstrates the complementary regulatory roles of inertia and time delay in chimera state formation. The temporal characteristics of $g_0(t)$ oscillations, shown in Fig.~\ref{fig8:heatmapchimera}~(b), provide further classification criteria. Regular periodic oscillations signify breathing chimera states, whereas irregular, chaotic fluctuations indicate turbulent chimeras.

In our final investigation of chimera states, Fig.~\ref {fig10:chimera} presents the temporal evolution of local order parameters, oscillator frequencies, $g_0(t)$ dynamics, $h_0$ values, and phase distributions across varying inertia ($m$) and time delay ($\tau$) parameters. The persistent condition $0 < g_0(t) < 1$ for all $t$ demonstrates the effective induction of chimera states through time delays, while the irregular fluctuations in $g_0(t)$ reveal inherent turbulence within these states.

For $m=0$ (panels (a) and (b)), spatial coherence falls to a value nearly zero, bringing chimera states close to disappearance. The introduction of inertia ($m \neq 0$) elevates $g_0(t)$ values and expands the spatial extent of chimera states, as evidenced in panels (c) and (d) for $m=5$, without altering $h_0$ values. This preservation of $h_0 \approx 0$ across all cases confirms the moving nature of these turbulent chimera states.

Frequency analysis reveals two distinct dynamical regimes: panels (a) and (c) exhibit low-frequency synchronized clusters around zero, while panels (b) and (d) show higher frequencies marking the transition from one area to another. This progressive synchronization pattern propagates through the rotor network in a manner reminiscent of partial epileptic seizure dynamics in neural systems. The observed behavior, consistent with previous findings in delayed coupled oscillator systems \cite{mahdavi2025synchronization}, emerges specifically in the right zone (large $\tau$ values).

\section{Conclusion\label{conclusion}}

In conclusion, we have investigated the effects of time delay in the presence of inertia within the Kuramoto model on a regular network of identical rotors. We first analyzed the master stability function and characteristic equation in the context of time-delayed interactions. Analytical results revealed the emergence of multistability as the time delay increased. Incorporating inertia led to the destabilization and elimination of certain phase-locked solutions. Notably, the relationship between increased inertia and the instability of phase-locked solutions seemed ambiguous. While it generally destabilized some solutions, in specific cases, it could also restore stability to previously unstable states.

Numerical computations indicate that the basin of attraction of phase-locked solutions closely aligned with the intrinsic frequencies of the rotors is significantly larger~\cite{mahdavi2025synchronization}. Furthermore, as the time delay extends, we encounter intervals where the likelihood of observing these solutions diminishes to zero, even though phase-locked solutions exist. This interval expands in correlation with the increasing inertia, suggesting a reduction in the basin of attraction for phase-locked solutions with increased inertia. A frequency distribution emerges where the probability function scarcely converges towards the Dirac delta function within these intervals. Moreover, in regions where the linear phase-locked solution loses stability but contains a larger basin of attraction, travelling wave solutions emerge,  propagating coherently across the ring.
These solutions, having asymmetric phase-lag configurations, demonstrate the capability to attain an order parameter between zero and one, which is constant over time. 

Finally, we have examined the chimera states. Our findings indicated that time delay can induce chimera states within the regular ring network. The irregular behavior of the spatial coherence function and the large fraction of strongly correlated nodes characterize these chimera states as moving turbulent. Additionally, for any value of inertia, chimera states manifest in two distinct regions. With increasing inertia, in the first region, with a smaller time delay, and in the second region, with a larger time delay, chimera states become apparent. Moreover, the size of chimera states expands in the presence of inertia. Also, in certain chimera states, a phenomenon is noted where multiple rotors synchronize together for a while, and then this behavior is transferred to other rotors with high angular velocities. This characteristic bears a resemblance to the temporal dynamics observed in partial epileptic seizures.

\section*{Funding}
P.H. acknowledges support by the Deutsche Forschungsgemeinschaft (DFG, German Research Foundation) – Project-ID 434434223 – SFB 1461. 

\section*{Declaration of competing interest}
The authors declare that they have no known competing financial interests or personal relationships that could have appeared to influence the work reported in this paper.

\section*{Data Availability Statement}
Data sharing is not applicable to this article as no datasets were generated or analyzed during the current study. The complete set of parameters and equations can be found in the model and methods section.


\begin{thebibliography}{34}%
\makeatletter
\providecommand \@ifxundefined [1]{%
 \@ifx{#1\undefined}
}%
\providecommand \@ifnum [1]{%
 \ifnum #1\expandafter \@firstoftwo
 \else \expandafter \@secondoftwo
 \fi
}%
\providecommand \@ifx [1]{%
 \ifx #1\expandafter \@firstoftwo
 \else \expandafter \@secondoftwo
 \fi
}%
\providecommand \natexlab [1]{#1}%
\providecommand \enquote  [1]{``#1''}%
\providecommand \bibnamefont  [1]{#1}%
\providecommand \bibfnamefont [1]{#1}%
\providecommand \citenamefont [1]{#1}%
\providecommand \href@noop [0]{\@secondoftwo}%
\providecommand \href [0]{\begingroup \@sanitize@url \@href}%
\providecommand \@href[1]{\@@startlink{#1}\@@href}%
\providecommand \@@href[1]{\endgroup#1\@@endlink}%
\providecommand \@sanitize@url [0]{\catcode `\\12\catcode `\$12\catcode
  `\&12\catcode `\#12\catcode `\^12\catcode `\_12\catcode `\%12\relax}%
\providecommand \@@startlink[1]{}%
\providecommand \@@endlink[0]{}%
\providecommand \url  [0]{\begingroup\@sanitize@url \@url }%
\providecommand \@url [1]{\endgroup\@href {#1}{\urlprefix }}%
\providecommand \urlprefix  [0]{URL }%
\providecommand \Eprint [0]{\href }%
\providecommand \doibase [0]{http://dx.doi.org/}%
\providecommand \selectlanguage [0]{\@gobble}%
\providecommand \bibinfo  [0]{\@secondoftwo}%
\providecommand \bibfield  [0]{\@secondoftwo}%
\providecommand \translation [1]{[#1]}%
\providecommand \BibitemOpen [0]{}%
\providecommand \bibitemStop [0]{}%
\providecommand \bibitemNoStop [0]{.\EOS\space}%
\providecommand \EOS [0]{\spacefactor3000\relax}%
\providecommand \BibitemShut  [1]{\csname bibitem#1\endcsname}%
\let\auto@bib@innerbib\@empty
\bibitem [{\citenamefont {Kuramoto}(1975)}]{kuramoto1975self}%
  \BibitemOpen
  \bibfield  {author} {\bibinfo {author} {\bibfnamefont {Y.}~\bibnamefont
  {Kuramoto}},\ }\bibfield  {title} {\enquote {\bibinfo {title}
  {Self-entrainment of a population of coupled non-linear oscillators},}\ }in\
  \href@noop {} {\emph {\bibinfo {booktitle} {International symposium on
  mathematical problems in theoretical physics}}}\ (\bibinfo {organization}
  {Springer},\ \bibinfo {year} {1975})\ pp.\ \bibinfo {pages}
  {420--422}\BibitemShut {NoStop}%
\bibitem [{\citenamefont {Tyson}(1973)}]{tyson1973some}%
  \BibitemOpen
  \bibfield  {author} {\bibinfo {author} {\bibfnamefont {J.~J.}\ \bibnamefont
  {Tyson}},\ }\bibfield  {title} {\enquote {\bibinfo {title} {Some further
  studies of nonlinear oscillations in chemical systems},}\ }\href@noop {}
  {\bibfield  {journal} {\bibinfo  {journal} {The Journal of Chemical Physics}\
  }\textbf {\bibinfo {volume} {58}},\ \bibinfo {pages} {3919--3930} (\bibinfo
  {year} {1973})}\BibitemShut {NoStop}%
\bibitem [{\citenamefont {Mirollo}\ and\ \citenamefont
  {Strogatz}(1990)}]{mirollo1990synchronization}%
  \BibitemOpen
  \bibfield  {author} {\bibinfo {author} {\bibfnamefont {R.~E.}\ \bibnamefont
  {Mirollo}}\ and\ \bibinfo {author} {\bibfnamefont {S.~H.}\ \bibnamefont
  {Strogatz}},\ }\bibfield  {title} {\enquote {\bibinfo {title}
  {Synchronization of pulse-coupled biological oscillators},}\ }\href@noop {}
  {\bibfield  {journal} {\bibinfo  {journal} {SIAM Journal on Applied
  Mathematics}\ }\textbf {\bibinfo {volume} {50}},\ \bibinfo {pages}
  {1645--1662} (\bibinfo {year} {1990})}\BibitemShut {NoStop}%
\bibitem [{\citenamefont {Nijmeijer}\ and\ \citenamefont
  {Rodriguez-Angeles}(2003)}]{nijmeijer2003synchronization}%
  \BibitemOpen
  \bibfield  {author} {\bibinfo {author} {\bibfnamefont {H.}~\bibnamefont
  {Nijmeijer}}\ and\ \bibinfo {author} {\bibfnamefont {A.}~\bibnamefont
  {Rodriguez-Angeles}},\ }\href@noop {} {\emph {\bibinfo {title}
  {Synchronization of mechanical systems}}},\ Vol.~\bibinfo {volume} {46}\
  (\bibinfo  {publisher} {World Scientific},\ \bibinfo {year}
  {2003})\BibitemShut {NoStop}%
\bibitem [{\citenamefont {Ermentrout}(1991)}]{ermentrout1991adaptive}%
  \BibitemOpen
  \bibfield  {author} {\bibinfo {author} {\bibfnamefont {B.}~\bibnamefont
  {Ermentrout}},\ }\bibfield  {title} {\enquote {\bibinfo {title} {An adaptive
  model for synchrony in the firefly pteroptyx malaccae},}\ }\href@noop {}
  {\bibfield  {journal} {\bibinfo  {journal} {Journal of Mathematical Biology}\
  }\textbf {\bibinfo {volume} {29}},\ \bibinfo {pages} {571--585} (\bibinfo
  {year} {1991})}\BibitemShut {NoStop}%
\bibitem [{\citenamefont {Tanaka}, \citenamefont {Lichtenberg},\ and\
  \citenamefont {Oishi}(1997{\natexlab{a}})}]{tanaka1997self}%
  \BibitemOpen
  \bibfield  {author} {\bibinfo {author} {\bibfnamefont {H.-A.}\ \bibnamefont
  {Tanaka}}, \bibinfo {author} {\bibfnamefont {A.~J.}\ \bibnamefont
  {Lichtenberg}}, \ and\ \bibinfo {author} {\bibfnamefont {S.}~\bibnamefont
  {Oishi}},\ }\bibfield  {title} {\enquote {\bibinfo {title}
  {Self-synchronization of coupled oscillators with hysteretic responses},}\
  }\href@noop {} {\bibfield  {journal} {\bibinfo  {journal} {Physica D:
  Nonlinear Phenomena}\ }\textbf {\bibinfo {volume} {100}},\ \bibinfo {pages}
  {279--300} (\bibinfo {year} {1997}{\natexlab{a}})}\BibitemShut {NoStop}%
\bibitem [{\citenamefont {Tanaka}, \citenamefont {Lichtenberg},\ and\
  \citenamefont {Oishi}(1997{\natexlab{b}})}]{tanaka1997first}%
  \BibitemOpen
  \bibfield  {author} {\bibinfo {author} {\bibfnamefont {H.-A.}\ \bibnamefont
  {Tanaka}}, \bibinfo {author} {\bibfnamefont {A.~J.}\ \bibnamefont
  {Lichtenberg}}, \ and\ \bibinfo {author} {\bibfnamefont {S.}~\bibnamefont
  {Oishi}},\ }\bibfield  {title} {\enquote {\bibinfo {title} {First order phase
  transition resulting from finite inertia in coupled oscillator systems},}\
  }\href@noop {} {\bibfield  {journal} {\bibinfo  {journal} {Physical Review
  Letters}\ }\textbf {\bibinfo {volume} {78}},\ \bibinfo {pages} {2104}
  (\bibinfo {year} {1997}{\natexlab{b}})}\BibitemShut {NoStop}%
\bibitem [{\citenamefont {Trees}, \citenamefont {Saranathan},\ and\
  \citenamefont {Stroud}(2005)}]{trees2005synchronization}%
  \BibitemOpen
  \bibfield  {author} {\bibinfo {author} {\bibfnamefont {B.~R.}\ \bibnamefont
  {Trees}}, \bibinfo {author} {\bibfnamefont {V.}~\bibnamefont {Saranathan}}, \
  and\ \bibinfo {author} {\bibfnamefont {D.}~\bibnamefont {Stroud}},\
  }\bibfield  {title} {\enquote {\bibinfo {title} {Synchronization in
  disordered josephson junction arrays: Small-world connections and the
  kuramoto model},}\ }\href@noop {} {\bibfield  {journal} {\bibinfo  {journal}
  {Physical Review E}\ }\textbf {\bibinfo {volume} {71}},\ \bibinfo {pages}
  {016215} (\bibinfo {year} {2005})}\BibitemShut {NoStop}%
\bibitem [{\citenamefont {Filatrella}, \citenamefont {Nielsen},\ and\
  \citenamefont {Pedersen}(2008)}]{filatrella2008analysis}%
  \BibitemOpen
  \bibfield  {author} {\bibinfo {author} {\bibfnamefont {G.}~\bibnamefont
  {Filatrella}}, \bibinfo {author} {\bibfnamefont {A.~H.}\ \bibnamefont
  {Nielsen}}, \ and\ \bibinfo {author} {\bibfnamefont {N.~F.}\ \bibnamefont
  {Pedersen}},\ }\bibfield  {title} {\enquote {\bibinfo {title} {Analysis of a
  power grid using a kuramoto-like model},}\ }\href@noop {} {\bibfield
  {journal} {\bibinfo  {journal} {The European Physical Journal B}\ }\textbf
  {\bibinfo {volume} {61}},\ \bibinfo {pages} {485--491} (\bibinfo {year}
  {2008})}\BibitemShut {NoStop}%
\bibitem [{\citenamefont {Rohden}\ \emph {et~al.}(2012)\citenamefont {Rohden},
  \citenamefont {Sorge}, \citenamefont {Timme},\ and\ \citenamefont
  {Witthaut}}]{rohden2012self}%
  \BibitemOpen
  \bibfield  {author} {\bibinfo {author} {\bibfnamefont {M.}~\bibnamefont
  {Rohden}}, \bibinfo {author} {\bibfnamefont {A.}~\bibnamefont {Sorge}},
  \bibinfo {author} {\bibfnamefont {M.}~\bibnamefont {Timme}}, \ and\ \bibinfo
  {author} {\bibfnamefont {D.}~\bibnamefont {Witthaut}},\ }\bibfield  {title}
  {\enquote {\bibinfo {title} {Self-organized synchronization in decentralized
  power grids},}\ }\href@noop {} {\bibfield  {journal} {\bibinfo  {journal}
  {Physical Review Letters}\ }\textbf {\bibinfo {volume} {109}},\ \bibinfo
  {pages} {064101} (\bibinfo {year} {2012})}\BibitemShut {NoStop}%
\bibitem [{\citenamefont {Rohden}\ \emph {et~al.}(2014)\citenamefont {Rohden},
  \citenamefont {Sorge}, \citenamefont {Witthaut},\ and\ \citenamefont
  {Timme}}]{rohden2014impact}%
  \BibitemOpen
  \bibfield  {author} {\bibinfo {author} {\bibfnamefont {M.}~\bibnamefont
  {Rohden}}, \bibinfo {author} {\bibfnamefont {A.}~\bibnamefont {Sorge}},
  \bibinfo {author} {\bibfnamefont {D.}~\bibnamefont {Witthaut}}, \ and\
  \bibinfo {author} {\bibfnamefont {M.}~\bibnamefont {Timme}},\ }\bibfield
  {title} {\enquote {\bibinfo {title} {Impact of network topology on synchrony
  of oscillatory power grids},}\ }\href@noop {} {\bibfield  {journal} {\bibinfo
   {journal} {Chaos: An Interdisciplinary Journal of Nonlinear Science}\
  }\textbf {\bibinfo {volume} {24}},\ \bibinfo {pages} {013123} (\bibinfo
  {year} {2014})}\BibitemShut {NoStop}%
\bibitem [{\citenamefont {Grzybowski}, \citenamefont {Macau},\ and\
  \citenamefont {Yoneyama}(2016)}]{grzybowski2016synchronization}%
  \BibitemOpen
  \bibfield  {author} {\bibinfo {author} {\bibfnamefont {J.~M.~V.}\
  \bibnamefont {Grzybowski}}, \bibinfo {author} {\bibfnamefont {E.~E.~N.}\
  \bibnamefont {Macau}}, \ and\ \bibinfo {author} {\bibfnamefont
  {T.}~\bibnamefont {Yoneyama}},\ }\bibfield  {title} {\enquote {\bibinfo
  {title} {On synchronization in power-grids modelled as networks of
  second-order kuramoto oscillators},}\ }\href@noop {} {\bibfield  {journal}
  {\bibinfo  {journal} {Chaos: An Interdisciplinary Journal of Nonlinear
  Science}\ }\textbf {\bibinfo {volume} {26}},\ \bibinfo {pages} {113113}
  (\bibinfo {year} {2016})}\BibitemShut {NoStop}%
\bibitem [{\citenamefont {Dolan}, \citenamefont {Majtanik},\ and\ \citenamefont
  {Tass}(2005)}]{dolan2005phase}%
  \BibitemOpen
  \bibfield  {author} {\bibinfo {author} {\bibfnamefont {K.}~\bibnamefont
  {Dolan}}, \bibinfo {author} {\bibfnamefont {M.}~\bibnamefont {Majtanik}}, \
  and\ \bibinfo {author} {\bibfnamefont {P.~A.}\ \bibnamefont {Tass}},\
  }\bibfield  {title} {\enquote {\bibinfo {title} {Phase resetting and
  transient desynchronization in networks of globally coupled phase oscillators
  with inertia},}\ }\href@noop {} {\bibfield  {journal} {\bibinfo  {journal}
  {Physica D: Nonlinear Phenomena}\ }\textbf {\bibinfo {volume} {211}},\
  \bibinfo {pages} {128--138} (\bibinfo {year} {2005})}\BibitemShut {NoStop}%
\bibitem [{\citenamefont {Majtanik}, \citenamefont {Dolan},\ and\ \citenamefont
  {Tass}(2006)}]{majtanik2006desynchronization}%
  \BibitemOpen
  \bibfield  {author} {\bibinfo {author} {\bibfnamefont {M.}~\bibnamefont
  {Majtanik}}, \bibinfo {author} {\bibfnamefont {K.}~\bibnamefont {Dolan}}, \
  and\ \bibinfo {author} {\bibfnamefont {P.~A.}\ \bibnamefont {Tass}},\
  }\bibfield  {title} {\enquote {\bibinfo {title} {Desynchronization in
  networks of globally coupled neurons with dendritic dynamics},}\ }\href@noop
  {} {\bibfield  {journal} {\bibinfo  {journal} {Journal of Biological
  Physics}\ }\textbf {\bibinfo {volume} {32}},\ \bibinfo {pages} {307--333}
  (\bibinfo {year} {2006})}\BibitemShut {NoStop}%
\bibitem [{\citenamefont {Sakyte}\ and\ \citenamefont
  {Ragulskis}(2011)}]{sakyte2011self}%
  \BibitemOpen
  \bibfield  {author} {\bibinfo {author} {\bibfnamefont {E.}~\bibnamefont
  {Sakyte}}\ and\ \bibinfo {author} {\bibfnamefont {M.}~\bibnamefont
  {Ragulskis}},\ }\bibfield  {title} {\enquote {\bibinfo {title} {Self-calming
  of a random network of dendritic neurons},}\ }\href@noop {} {\bibfield
  {journal} {\bibinfo  {journal} {Neurocomputing}\ }\textbf {\bibinfo {volume}
  {74}},\ \bibinfo {pages} {3912--3920} (\bibinfo {year} {2011})}\BibitemShut
  {NoStop}%
\bibitem [{\citenamefont {Mishra}\ \emph {et~al.}(2021)\citenamefont {Mishra},
  \citenamefont {Ghosh}, \citenamefont {Kumar~Dana}, \citenamefont
  {Kapitaniak},\ and\ \citenamefont {Hens}}]{mishra2021neuron}%
  \BibitemOpen
  \bibfield  {author} {\bibinfo {author} {\bibfnamefont {A.}~\bibnamefont
  {Mishra}}, \bibinfo {author} {\bibfnamefont {S.}~\bibnamefont {Ghosh}},
  \bibinfo {author} {\bibfnamefont {S.}~\bibnamefont {Kumar~Dana}}, \bibinfo
  {author} {\bibfnamefont {T.}~\bibnamefont {Kapitaniak}}, \ and\ \bibinfo
  {author} {\bibfnamefont {C.}~\bibnamefont {Hens}},\ }\bibfield  {title}
  {\enquote {\bibinfo {title} {Neuron-like spiking and bursting in josephson
  junctions: A review},}\ }\href@noop {} {\bibfield  {journal} {\bibinfo
  {journal} {Chaos: An Interdisciplinary Journal of Nonlinear Science}\
  }\textbf {\bibinfo {volume} {31}},\ \bibinfo {pages} {052101} (\bibinfo
  {year} {2021})}\BibitemShut {NoStop}%
\bibitem [{\citenamefont {Kozyreff}, \citenamefont {Vladimirov},\ and\
  \citenamefont {Mandel}(2000)}]{kozyreff2000global}%
  \BibitemOpen
  \bibfield  {author} {\bibinfo {author} {\bibfnamefont {G.}~\bibnamefont
  {Kozyreff}}, \bibinfo {author} {\bibfnamefont {A.}~\bibnamefont
  {Vladimirov}}, \ and\ \bibinfo {author} {\bibfnamefont {P.}~\bibnamefont
  {Mandel}},\ }\bibfield  {title} {\enquote {\bibinfo {title} {Global coupling
  with time delay in an array of semiconductor lasers},}\ }\href@noop {}
  {\bibfield  {journal} {\bibinfo  {journal} {Physical Review Letters}\
  }\textbf {\bibinfo {volume} {85}},\ \bibinfo {pages} {3809} (\bibinfo {year}
  {2000})}\BibitemShut {NoStop}%
\bibitem [{\citenamefont {Reddy}, \citenamefont {Sen},\ and\ \citenamefont
  {Johnston}(2000)}]{reddy2000experimental}%
  \BibitemOpen
  \bibfield  {author} {\bibinfo {author} {\bibfnamefont {D.~R.}\ \bibnamefont
  {Reddy}}, \bibinfo {author} {\bibfnamefont {A.}~\bibnamefont {Sen}}, \ and\
  \bibinfo {author} {\bibfnamefont {G.~L.}\ \bibnamefont {Johnston}},\
  }\bibfield  {title} {\enquote {\bibinfo {title} {Experimental evidence of
  time-delay-induced death in coupled limit-cycle oscillators},}\ }\href@noop
  {} {\bibfield  {journal} {\bibinfo  {journal} {Physical Review Letters}\
  }\textbf {\bibinfo {volume} {85}},\ \bibinfo {pages} {3381} (\bibinfo {year}
  {2000})}\BibitemShut {NoStop}%
\bibitem [{\citenamefont {Yeung}\ and\ \citenamefont
  {Strogatz}(1999)}]{yeung1999time}%
  \BibitemOpen
  \bibfield  {author} {\bibinfo {author} {\bibfnamefont {M.~S.}\ \bibnamefont
  {Yeung}}\ and\ \bibinfo {author} {\bibfnamefont {S.~H.}\ \bibnamefont
  {Strogatz}},\ }\bibfield  {title} {\enquote {\bibinfo {title} {Time delay in
  the kuramoto model of coupled oscillators},}\ }\href@noop {} {\bibfield
  {journal} {\bibinfo  {journal} {Physical Review Letters}\ }\textbf {\bibinfo
  {volume} {82}},\ \bibinfo {pages} {648} (\bibinfo {year} {1999})}\BibitemShut
  {NoStop}%
\bibitem [{\citenamefont {Kerszberg}\ and\ \citenamefont
  {Zippelius}(1990)}]{kerszberg1990synchronization}%
  \BibitemOpen
  \bibfield  {author} {\bibinfo {author} {\bibfnamefont {M.}~\bibnamefont
  {Kerszberg}}\ and\ \bibinfo {author} {\bibfnamefont {A.}~\bibnamefont
  {Zippelius}},\ }\bibfield  {title} {\enquote {\bibinfo {title}
  {Synchronization in neural assemblies},}\ }\href@noop {} {\bibfield
  {journal} {\bibinfo  {journal} {Physica Scripta}\ }\textbf {\bibinfo {volume}
  {1990}},\ \bibinfo {pages} {54} (\bibinfo {year} {1990})}\BibitemShut
  {NoStop}%
\bibitem [{\citenamefont {Waibel}\ \emph {et~al.}(2013)\citenamefont {Waibel},
  \citenamefont {Hanazawa}, \citenamefont {Hinton}, \citenamefont {Shikano},\
  and\ \citenamefont {Lang}}]{waibel2013phoneme}%
  \BibitemOpen
  \bibfield  {author} {\bibinfo {author} {\bibfnamefont {A.}~\bibnamefont
  {Waibel}}, \bibinfo {author} {\bibfnamefont {T.}~\bibnamefont {Hanazawa}},
  \bibinfo {author} {\bibfnamefont {G.}~\bibnamefont {Hinton}}, \bibinfo
  {author} {\bibfnamefont {K.}~\bibnamefont {Shikano}}, \ and\ \bibinfo
  {author} {\bibfnamefont {K.~J.}\ \bibnamefont {Lang}},\ }\bibfield  {title}
  {\enquote {\bibinfo {title} {Phoneme recognition using time-delay neural
  networks},}\ }in\ \href@noop {} {\emph {\bibinfo {booktitle}
  {Backpropagation}}}\ (\bibinfo  {publisher} {Psychology Press},\ \bibinfo
  {year} {2013})\ pp.\ \bibinfo {pages} {35--61}\BibitemShut {NoStop}%
\bibitem [{\citenamefont {Schuster}\ and\ \citenamefont
  {Wagner}(1989)}]{schuster1989mutual}%
  \BibitemOpen
  \bibfield  {author} {\bibinfo {author} {\bibfnamefont {H.~G.}\ \bibnamefont
  {Schuster}}\ and\ \bibinfo {author} {\bibfnamefont {P.}~\bibnamefont
  {Wagner}},\ }\bibfield  {title} {\enquote {\bibinfo {title} {Mutual
  entrainment of two limit cycle oscillators with time delayed coupling},}\
  }\href@noop {} {\bibfield  {journal} {\bibinfo  {journal} {Progress of
  Theoretical Physics}\ }\textbf {\bibinfo {volume} {81}},\ \bibinfo {pages}
  {939--945} (\bibinfo {year} {1989})}\BibitemShut {NoStop}%
\bibitem [{\citenamefont {Mahdavi}, \citenamefont {Zarei},\ and\ \citenamefont
  {Shahbazi}(2025)}]{mahdavi2025synchronization}%
  \BibitemOpen
  \bibfield  {author} {\bibinfo {author} {\bibfnamefont {E.}~\bibnamefont
  {Mahdavi}}, \bibinfo {author} {\bibfnamefont {M.}~\bibnamefont {Zarei}}, \
  and\ \bibinfo {author} {\bibfnamefont {F.}~\bibnamefont {Shahbazi}},\
  }\bibfield  {title} {\enquote {\bibinfo {title} {Synchronization of two
  coupled massive oscillators in the time-delayed kuramoto model},}\
  }\href@noop {} {\bibfield  {journal} {\bibinfo  {journal} {Chaos: An
  Interdisciplinary Journal of Nonlinear Science}\ }\textbf {\bibinfo {volume}
  {35}},\ \bibinfo {pages} {013122} (\bibinfo {year} {2025})}\BibitemShut
  {NoStop}%
\bibitem [{\citenamefont {Choi}\ \emph {et~al.}(2000)\citenamefont {Choi},
  \citenamefont {Kim}, \citenamefont {Kim},\ and\ \citenamefont
  {Hong}}]{choi2000synchronization}%
  \BibitemOpen
  \bibfield  {author} {\bibinfo {author} {\bibfnamefont {M.}~\bibnamefont
  {Choi}}, \bibinfo {author} {\bibfnamefont {H.}~\bibnamefont {Kim}}, \bibinfo
  {author} {\bibfnamefont {D.}~\bibnamefont {Kim}}, \ and\ \bibinfo {author}
  {\bibfnamefont {H.}~\bibnamefont {Hong}},\ }\bibfield  {title} {\enquote
  {\bibinfo {title} {Synchronization in a system of globally coupled
  oscillators with time delay},}\ }\href@noop {} {\bibfield  {journal}
  {\bibinfo  {journal} {Physical Review E}\ }\textbf {\bibinfo {volume} {61}},\
  \bibinfo {pages} {371} (\bibinfo {year} {2000})}\BibitemShut {NoStop}%
\bibitem [{\citenamefont {Madadi~Asl}, \citenamefont {Valizadeh},\ and\
  \citenamefont {Tass}(2018)}]{madadi2018delay}%
  \BibitemOpen
  \bibfield  {author} {\bibinfo {author} {\bibfnamefont {M.}~\bibnamefont
  {Madadi~Asl}}, \bibinfo {author} {\bibfnamefont {A.}~\bibnamefont
  {Valizadeh}}, \ and\ \bibinfo {author} {\bibfnamefont {P.~A.}\ \bibnamefont
  {Tass}},\ }\bibfield  {title} {\enquote {\bibinfo {title} {Delay-induced
  multistability and loop formation in neuronal networks with
  spike-timing-dependent plasticity},}\ }\href@noop {} {\bibfield  {journal}
  {\bibinfo  {journal} {Scientific Reports}\ }\textbf {\bibinfo {volume} {8}},\
  \bibinfo {pages} {12068} (\bibinfo {year} {2018})}\BibitemShut {NoStop}%
\bibitem [{\citenamefont {Ameli}, \citenamefont {Karimian},\ and\ \citenamefont
  {Shahbazi}(2021)}]{ameli2021time}%
  \BibitemOpen
  \bibfield  {author} {\bibinfo {author} {\bibfnamefont {S.}~\bibnamefont
  {Ameli}}, \bibinfo {author} {\bibfnamefont {M.}~\bibnamefont {Karimian}}, \
  and\ \bibinfo {author} {\bibfnamefont {F.}~\bibnamefont {Shahbazi}},\
  }\bibfield  {title} {\enquote {\bibinfo {title} {Time-delayed kuramoto model
  in the watts--strogatz small-world networks},}\ }\href@noop {} {\bibfield
  {journal} {\bibinfo  {journal} {Chaos: An Interdisciplinary Journal of
  Nonlinear Science}\ }\textbf {\bibinfo {volume} {31}},\ \bibinfo {pages}
  {113125} (\bibinfo {year} {2021})}\BibitemShut {NoStop}%
\bibitem [{\citenamefont {Vicente}\ \emph {et~al.}(2008)\citenamefont
  {Vicente}, \citenamefont {Gollo}, \citenamefont {Mirasso}, \citenamefont
  {Fischer},\ and\ \citenamefont {Pipa}}]{vicente2008dynamical}%
  \BibitemOpen
  \bibfield  {author} {\bibinfo {author} {\bibfnamefont {R.}~\bibnamefont
  {Vicente}}, \bibinfo {author} {\bibfnamefont {L.~L.}\ \bibnamefont {Gollo}},
  \bibinfo {author} {\bibfnamefont {C.~R.}\ \bibnamefont {Mirasso}}, \bibinfo
  {author} {\bibfnamefont {I.}~\bibnamefont {Fischer}}, \ and\ \bibinfo
  {author} {\bibfnamefont {G.}~\bibnamefont {Pipa}},\ }\bibfield  {title}
  {\enquote {\bibinfo {title} {Dynamical relaying can yield zero time lag
  neuronal synchrony despite long conduction delays},}\ }\href@noop {}
  {\bibfield  {journal} {\bibinfo  {journal} {Proceedings of the National
  Academy of Sciences}\ }\textbf {\bibinfo {volume} {105}},\ \bibinfo {pages}
  {17157--17162} (\bibinfo {year} {2008})}\BibitemShut {NoStop}%
\bibitem [{\citenamefont {Lehnert}\ \emph {et~al.}(2011)\citenamefont
  {Lehnert}, \citenamefont {Dahms}, \citenamefont {H{\"o}vel},\ and\
  \citenamefont {Sch{\"o}ll}}]{lehnert2011loss}%
  \BibitemOpen
  \bibfield  {author} {\bibinfo {author} {\bibfnamefont {J.}~\bibnamefont
  {Lehnert}}, \bibinfo {author} {\bibfnamefont {T.}~\bibnamefont {Dahms}},
  \bibinfo {author} {\bibfnamefont {P.}~\bibnamefont {H{\"o}vel}}, \ and\
  \bibinfo {author} {\bibfnamefont {E.}~\bibnamefont {Sch{\"o}ll}},\ }\bibfield
   {title} {\enquote {\bibinfo {title} {Loss of synchronization in complex
  neuronal networks with delay},}\ }\href@noop {} {\bibfield  {journal}
  {\bibinfo  {journal} {Europhysics Letters}\ }\textbf {\bibinfo {volume}
  {96}},\ \bibinfo {pages} {60013} (\bibinfo {year} {2011})}\BibitemShut
  {NoStop}%
\bibitem [{\citenamefont {M{\'e}tivier}, \citenamefont {Wetzel},\ and\
  \citenamefont {Gupta}(2020)}]{metivier2020onset}%
  \BibitemOpen
  \bibfield  {author} {\bibinfo {author} {\bibfnamefont {D.}~\bibnamefont
  {M{\'e}tivier}}, \bibinfo {author} {\bibfnamefont {L.}~\bibnamefont
  {Wetzel}}, \ and\ \bibinfo {author} {\bibfnamefont {S.}~\bibnamefont
  {Gupta}},\ }\bibfield  {title} {\enquote {\bibinfo {title} {Onset of
  synchronization in networks of second-order kuramoto oscillators with delayed
  coupling: Exact results and application to phase-locked loops},}\ }\href@noop
  {} {\bibfield  {journal} {\bibinfo  {journal} {Physical Review Research}\
  }\textbf {\bibinfo {volume} {2}},\ \bibinfo {pages} {023183} (\bibinfo {year}
  {2020})}\BibitemShut {NoStop}%
\bibitem [{\citenamefont {Hong}, \citenamefont {Jeon},\ and\ \citenamefont
  {Choi}(2002)}]{hong2002spontaneous}%
  \BibitemOpen
  \bibfield  {author} {\bibinfo {author} {\bibfnamefont {H.}~\bibnamefont
  {Hong}}, \bibinfo {author} {\bibfnamefont {G.~S.}\ \bibnamefont {Jeon}}, \
  and\ \bibinfo {author} {\bibfnamefont {M.}~\bibnamefont {Choi}},\ }\bibfield
  {title} {\enquote {\bibinfo {title} {Spontaneous phase oscillation induced by
  inertia and time delay},}\ }\href@noop {} {\bibfield  {journal} {\bibinfo
  {journal} {Physical Review E}\ }\textbf {\bibinfo {volume} {65}},\ \bibinfo
  {pages} {026208} (\bibinfo {year} {2002})}\BibitemShut {NoStop}%
\bibitem [{\citenamefont {Prousalis}\ and\ \citenamefont
  {Wetzel}(2022)}]{prousalis2022synchronization}%
  \BibitemOpen
  \bibfield  {author} {\bibinfo {author} {\bibfnamefont {D.}~\bibnamefont
  {Prousalis}}\ and\ \bibinfo {author} {\bibfnamefont {L.}~\bibnamefont
  {Wetzel}},\ }\bibfield  {title} {\enquote {\bibinfo {title} {Synchronization
  in the presence of time delays and inertia: Stability criteria},}\
  }\href@noop {} {\bibfield  {journal} {\bibinfo  {journal} {Physical Review
  E}\ }\textbf {\bibinfo {volume} {105}},\ \bibinfo {pages} {014210} (\bibinfo
  {year} {2022})}\BibitemShut {NoStop}%
\bibitem [{\citenamefont {Dai}\ \emph {et~al.}(2018)\citenamefont {Dai},
  \citenamefont {Zhou}, \citenamefont {Peron}, \citenamefont {Lin},\ and\
  \citenamefont {Ji}}]{dai2018interplay}%
  \BibitemOpen
  \bibfield  {author} {\bibinfo {author} {\bibfnamefont {F.}~\bibnamefont
  {Dai}}, \bibinfo {author} {\bibfnamefont {S.}~\bibnamefont {Zhou}}, \bibinfo
  {author} {\bibfnamefont {T.}~\bibnamefont {Peron}}, \bibinfo {author}
  {\bibfnamefont {W.}~\bibnamefont {Lin}}, \ and\ \bibinfo {author}
  {\bibfnamefont {P.}~\bibnamefont {Ji}},\ }\bibfield  {title} {\enquote
  {\bibinfo {title} {Interplay among inertia, time delay, and frustration on
  synchronization dynamics},}\ }\href@noop {} {\bibfield  {journal} {\bibinfo
  {journal} {Physical Review E}\ }\textbf {\bibinfo {volume} {98}},\ \bibinfo
  {pages} {052218} (\bibinfo {year} {2018})}\BibitemShut {NoStop}%
\bibitem [{\citenamefont {Ameli}\ \emph {et~al.}(2025)\citenamefont {Ameli},
  \citenamefont {Mahdavi}, \citenamefont {Zarei},\ and\ \citenamefont
  {Shahbazi}}]{ameli2025synchronization}%
  \BibitemOpen
  \bibfield  {author} {\bibinfo {author} {\bibfnamefont {S.}~\bibnamefont
  {Ameli}}, \bibinfo {author} {\bibfnamefont {E.}~\bibnamefont {Mahdavi}},
  \bibinfo {author} {\bibfnamefont {M.}~\bibnamefont {Zarei}}, \ and\ \bibinfo
  {author} {\bibfnamefont {F.}~\bibnamefont {Shahbazi}},\ }\bibfield  {title}
  {\enquote {\bibinfo {title} {Synchronization of the time-delayed kuramoto
  model in a regular network},}\ }\href@noop {} {\bibfield  {journal} {\bibinfo
   {journal} {The European Physical Journal Plus}\ }\textbf {\bibinfo {volume}
  {140}},\ \bibinfo {pages} {813} (\bibinfo {year} {2025})}\BibitemShut
  {NoStop}%
\bibitem [{\citenamefont {Kemeth}\ \emph {et~al.}(2016)\citenamefont {Kemeth},
  \citenamefont {Haugland}, \citenamefont {Schmidt}, \citenamefont
  {Kevrekidis},\ and\ \citenamefont {Krischer}}]{kemeth2016classification}%
  \BibitemOpen
  \bibfield  {author} {\bibinfo {author} {\bibfnamefont {F.~P.}\ \bibnamefont
  {Kemeth}}, \bibinfo {author} {\bibfnamefont {S.~W.}\ \bibnamefont
  {Haugland}}, \bibinfo {author} {\bibfnamefont {L.}~\bibnamefont {Schmidt}},
  \bibinfo {author} {\bibfnamefont {I.~G.}\ \bibnamefont {Kevrekidis}}, \ and\
  \bibinfo {author} {\bibfnamefont {K.}~\bibnamefont {Krischer}},\ }\bibfield
  {title} {\enquote {\bibinfo {title} {A classification scheme for chimera
  states},}\ }\href@noop {} {\bibfield  {journal} {\bibinfo  {journal} {Chaos:
  An Interdisciplinary Journal of Nonlinear Science}\ }\textbf {\bibinfo
  {volume} {26}},\ \bibinfo {pages} {094815} (\bibinfo {year}
  {2016})}\BibitemShut {NoStop}%
\end{thebibliography}
%
\end{document}